# Specifications, quality control, manufacturing, and testing of accelerator magnets

*D. Einfeld*
CELLS-ALBA, Barcelona, Spain

**Abstract**
The performance of the magnets plays an important role in the functioning of an accelerator. Most of the magnets are designed at the accelerator laboratory and built by industry. The link between the laboratory and the manufacturer is the contract containing the Technical Specifications of the magnets. For an overview of the contents of the Technical Specifications, the specifications for the magnets of ALBA (bending, quadrupole, and sextupole) are described in this paper. The basic rules of magnet design are reviewed in Appendix A.

## 1   Introduction

The so-called Technical Specifications have to be prepared for the Call for Tender process of the magnets (see Appendix B). This will be the dominant document of the contract with the manufacturer. The main topics covered by the Technical Specifications document include:

- Specifications
- Quality control
- Manufacturing
- Acceptance test

Accelerator magnets include:

- Normal-conducting magnets (bending, quadrupole, sextupole, corrector).
- Superconducting magnets.
- Magnets using permanent magnet material.
- Fast-pulsed magnets.
- Specialized magnets such as septa and kickers used for injection and extraction.

The scope of this paper will be limited to conventional room-temperature, iron-dominated accelerator magnets, mainly bending, quadrupole, sextupole and corrector (see Fig. 1). The main components of the different types of iron-dominated accelerator magnets are (see Fig. 2):

- Iron yoke
- Pole profile
- Coils
- Manifolds
- Sensors or thermocouples
- Supports

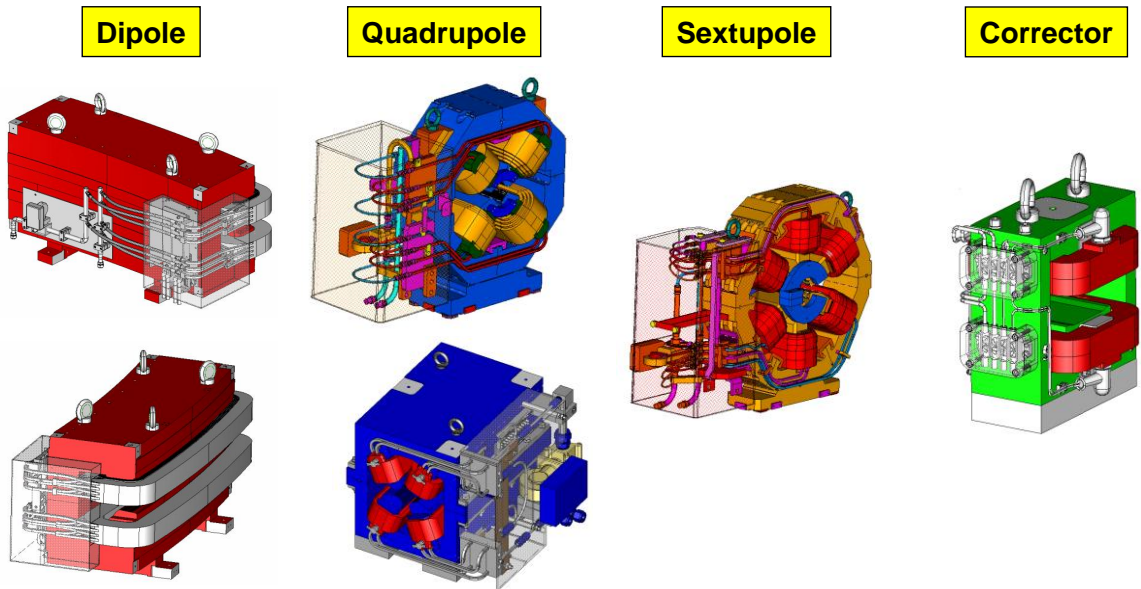

**Fig. 1:** Examples of different types of iron-dominated accelerator magnets

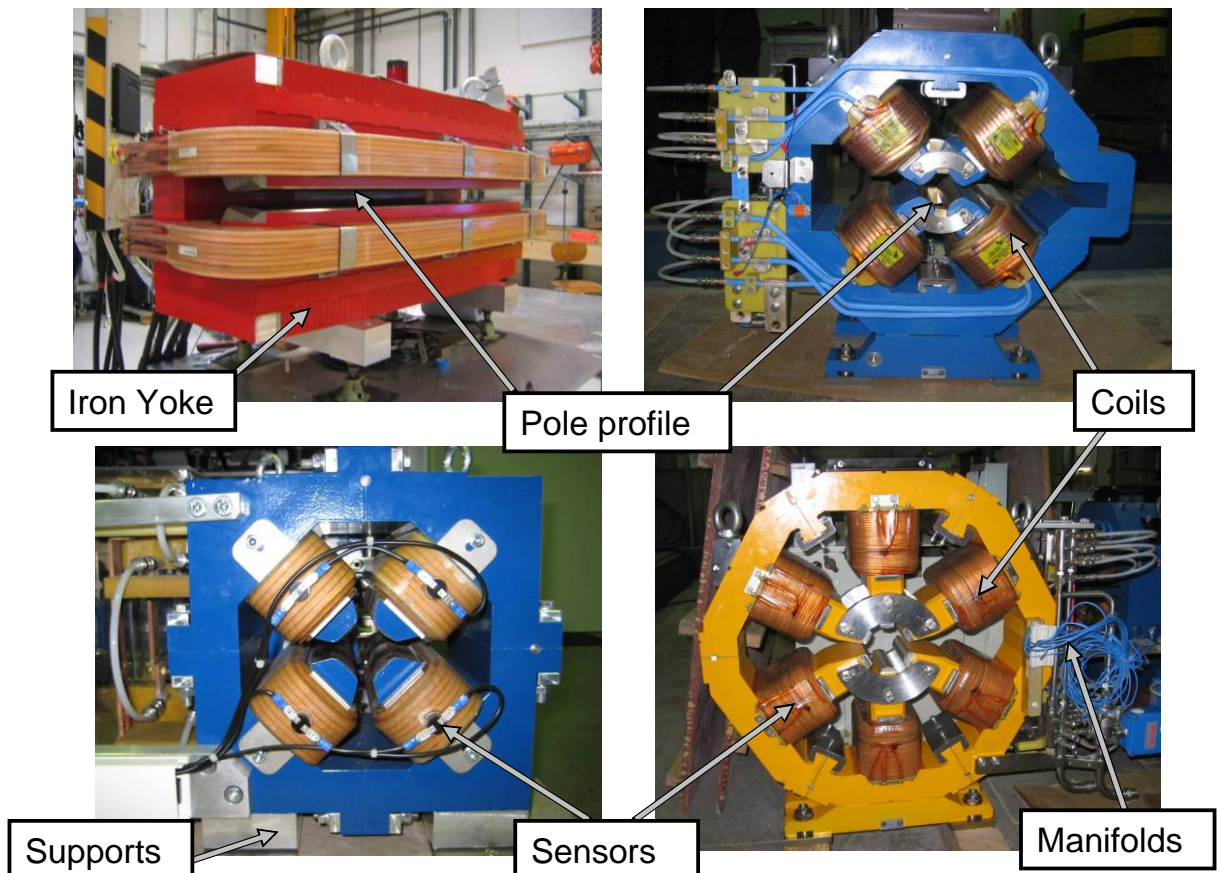

**Fig. 2:** Main components for different types of iron-dominated accelerator magnets

## 2 Contents of the Technical Specifications

A typical Technical Specifications document will have a structure similar to the following one:

2.1.) INTRODUCTION

2.2.) SCOPE OF CONTRACT

2.3.) BENDING MAGNET DETAILS

    2.3.1.) Description of the bending magnets

    2.3.2.) Parameter list

    2.3.3.) Specification drawings

    2.3.4.) Trim coils

    2.3.5.) Magnet support feet

    2.3.6.) Survey monument

    2.3.7.) Lifting brackets

2.4.) PHASING OF THE CONTRACT

2.5.) SCHEDULE

2.6.) TENDERING AND CONTRACT MANAGEMENT

    2.6.1.) Tendering

    2.6.2.) Information required with the tender

    2.6.3.) Contract management

    2.6.4.) Quality assurance

    2.6.5.) Documentation

    2.6.6.) Numbering

    2.6.7.) Guarantee

2.7.) MAGNETIC STEEL

    2.7.1.) Steel characteristics for the magnet

    2.7.2.) Laminations

    2.7.3.) Testing of steel

    2.7.4.) Steel supplier

2.8.) LAMINATIONS AND YOKE

    2.8.1.) Laminations stamping

    2.8.2.) Lamination stamping tests

    2.8.3.) Laminations shuffling

    2.8.4.) Yoke

    2.8.5.) Bending magnet ends

    2.8.6.) Mechanical yoke testing

    2.8.7.) Protection and painting

- 2.9.) COILS
    - 2.9.1.) Coils manufacturing
    - 2.9.2.) Conductor
    - 2.9.3.) Conductor supplier
    - 2.9.4.) Conductor tests before winding
    - 2.9.5.) Pancake winding
    - 2.9.6.) Pancake insulation and impregnation
    - 2.9.7.) Terminations of the coil
    - 2.9.8.) Coils testing
- 2.10.) MECHANICAL AND ELECTRICAL TESTS ON COMPLETE MAGNETS
    - 2.10.1.) Mechanical and electrical tests on complete pre-series magnet
    - 2.10.2.) Mechanical and electrical tests on complete production magnets
    - 2.10.3.) Acceptance tests after delivery (Site acceptance test)
- 2.11.) PACKING AND TRANSPORTATION
    - 2.11.1.) Packing
    - 2.11.2.) Transportation

Now we will proceed to examine in detail as an example the Technical Specifications document in the case of the bending magnets for the Storage Ring of ALBA. Some aspects regarding the specifications for the quadrupoles and the sextupoles will be described as well.

## 2.1 INTRODUCTION

- 2.1.1.) The Consortium for the Construction, Equipment and Exploitation of the Synchrotron Light Laboratory (CELLS) is responsible for the construction of a new synchrotron radiation facility, named ALBA. The facility will comprise a 3 GeV electron storage ring, injected from a 100 MeV Linac through a full energy booster synchrotron, and an initial complement of five beam lines.

- 2.1.2.) The Storage Ring is composed of 32 bending magnets, 112 quadrupole magnets and 120 sextupoles magnets, plus a given number of spare magnets.

- 2.1.3.) The present specification contains the technical specifications for the bending magnets. All dimensions and tolerances of the magnets are defined in the specification and the attached drawings. The small electron beam size of ALBA and its stability results in tight tolerances on the magnets.

- 2.1.4.) The contracts for the bending magnets will include the responsibility for the electrical, mechanical and thermal design of the magnets, their construction, followed by their mechanical and electrical testing; CELLS will retain responsibility for the magnetic field generated by the magnet poles and for the magnetic measurements. This specification therefore defines the required pole profile, together with the electrical and mechanical operational performance. The final design arrangements to achieve the specified features in a reproducible way in a correctly engineered magnet will be the responsibility of the manufacturer.

- 2.1.5.) The contracts for the quadrupoles will include the responsibility for the electrical, mechanical and thermal design of the magnets, their construction, followed by their mechanical and electrical testing, and the measurement of their magnetic performance; CELLS will retain responsibility for the magnetic field generated by the magnet poles. This specification therefore defines the required pole profile, together with the electrical and mechanical operational performance. The final design arrangements to achieve the specified features in a reproducible way in a correctly engineered magnet will be the responsibility of the manufacturer.

2.1.6.) It is desired to have full magnetic measuring of the production magnets carried out at the manufacturer under his responsibility. The manufacturer will provide the necessary magnetic field measuring equipment and suitable technical operators. After testing, it may be necessary to store magnets on the manufacturer's premises. Both these requirements are more fully described in this specification.

## 2.2 SCOPE OF CONTRACT

2.2.1.) This specification covers the engineering design, manufacture, assembly, testing and delivery of bending magnets for the Storage Ring of the ALBA facility.

2.2.2.) The specification also covers the supply of all materials and the construction of all tools, jigs and fixtures required to complete the contract.

2.2.3.) The magnets, as manufactured, will consist of a magnetic yoke, coils, and all other mechanical brackets and fittings required for their full assembly. They will also be fitted with support feet, water manifolds, electrical termination blocks, coil interconnections and insulated protective covers.

2.2.4.) There are not any items which have to be provided by CELLS.

## 2.3 MAGNET DETAILS

### *2.3.1 Description of the bending magnets*

2.3.1.1.) ALBA requires a total of 32+1 bending magnets.

2.3.1.2.) The bending magnets are curved parallel-ended C-type with a central gap of 36 mm.

2.3.1.3.) The magnets generate a magnetic field with a gradient ($G = 5.656$ T/m) and operate at a maximum induction of 1.42 T at the central point.

2.3.1.4.) The cores will be laminated and the laminations will be stacked together along a curved line with uniform radius thus forming a curved magnet with parallel ends or you have sector magnets.

2.3.1.5.) The bending magnets are required with an iron yoke length of 1340 mm.

2.3.1.6.) The location of the electron photon beam vacuum chamber in the gap of the magnets leads to the definition of a 'vacuum chamber stay clear area'. This is defined in the appropriate drawings. It will be part of the manufacturer's responsibility, during the mechanical and electrical design of the magnet, to keep this space clear for the vacuum components and to make allowance for this when designing the coil.

2.3.1.7.) The magnets will be excited by coils mounted on the poles. These coils will be wound from solid conductor with a central water cooling hole. The bending magnets will be powered by one power supply, all magnets are in series.

## 2.3.2 *Parameter list and drawings*

### 2.3.2.1 *Bending magnets*

Table 1 shows the list of specified parameters for the bending magnets of ALBA Storage Ring and Booster accelerators. Figures 3–7 show some of the design drawings for Storage Ring bending magnets.

**Table 1:** Parameter list of ALBA Storage Ring and Booster bending magnets

| Storage Ring Bending Magent | | | Booster Bending Magents | | | |
|---|---|---|---|---|---|---|
| | | | | | 5 Degr. | 10 Degr. |
| **Magnetic properties** | | | **Magnetic properties** | | | |
| Beam Energy (E) | GeV | 3 | Beam Energy (E) | GeV | 3 | 3 |
| Central Field (Bo) | T | 1.42 | Central Field (Bo) | T | 0.8733 | 0.8733 |
| Field gradient (Go) | T/m | 5.656 | Field gradient (Go) | T/m | 2.292 | 2.292 |
| Sextupole component (B") | T/(m^2) | 0 | Sextupole component (B") | T/(m^2) | 18 | 18 |
| Effective length (Lo) | m | 1.384 | Effective length (Lo) | m | 1 | 2 |
| **Mechanical properties** | | | **Mechanical properties** | | | |
| Bending radius (Ro) | m | 7.047 | Bending radius (Ro) | m | 11.4592 | 11.4592 |
| Bending angle (phi) | degrees | 11.25 | Bending angle (phi) | degrees | 5 | 10 |
| Central Gap (h) | mm | 36 | Central Gap (h) | mm | 22.6 | 22.6 |
| Length of Fe-yoke L(Fe) | mm | 1340 | Length of Fe-yoke L(Fe) | mm | 0.972 | 1.972 |
| **Coil and conductor** | | | **Coil and conductor** | | | |
| Number of coils | | 2 | Number of coils | | | 2 | 2 |
| Number of pancakes per coil | | 4 | Number of pancakes per coil | | 1 | 1 |
| Number of turns per pancake | | 10 | Number of turns per pancake | | 12 | 12 |
| Concductor size | mm^2 | 16.3*10.8 | Concductor size | mm^2 | 12*12 | 12*12 |
| Cooling channel diameter (D) | mm | 6.6 | Cooling channel diameter (D) | mm | 5 | 5 |
| Number of ampere turns per coil | A-turns | 20340 | Number of ampere turns per coil | A-turns | 7906 | 7906 |
| Current (I) | A | 509 | Current (I) | A | 659 | 659 |
| Current density (j) | A/mm^2 | 3.59 | Current density (j) | A/mm^2 | 6.08 | 6.08 |
| Resistance at 23 degrees | mΩ | 34.5 | Resistance at 23 degrees | mΩ | 9.2 | 18.2 |
| Inductivity | mH | 40 | Inductance | mH | 1.3 | 2.6 |
| Voltage drop | V | 17.6 | Voltage drop | V | 6.1 | 31.8 |
| Power | kW | 8.93 | Power | kW | 2 | 3.94 |
| **Cooling** | | | **Cooling** | | | |
| Maximim DT | Celsius | 8.6 | Maximim DT | Celsius | 11 | 11 |
| Nominal input temperature | Celsius | 23 | Nominal input temperature | Celsius | | 23 |
| Number of cooling circuits per coil | | 2 | Number of cooling circuits per coil | | 1 | 2 |
| Maximum pressure drop per magnet | bar | 7 | Maximum pressure drop per magne | bar | 7 | 7 |

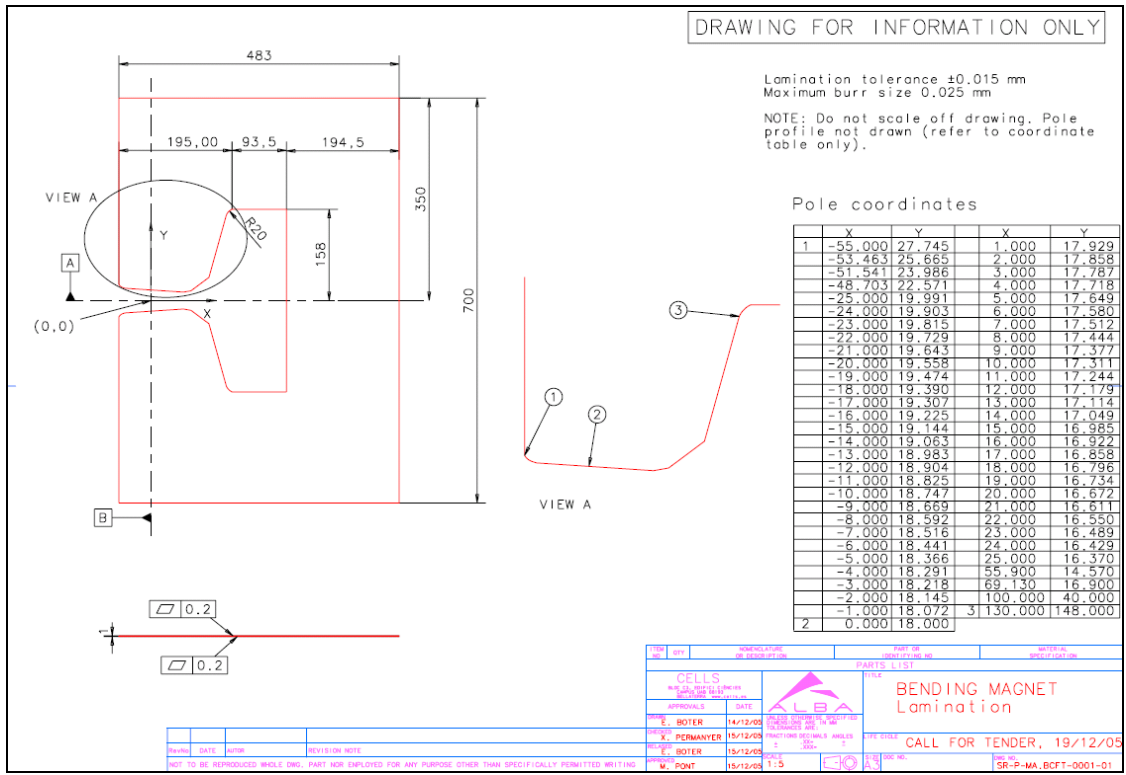

**Fig. 3:** Drawing showing the dimensions and the pole profile of the laminations for ALBA Storage Ring bending magnets. The required production tolerance is ±15 μm.

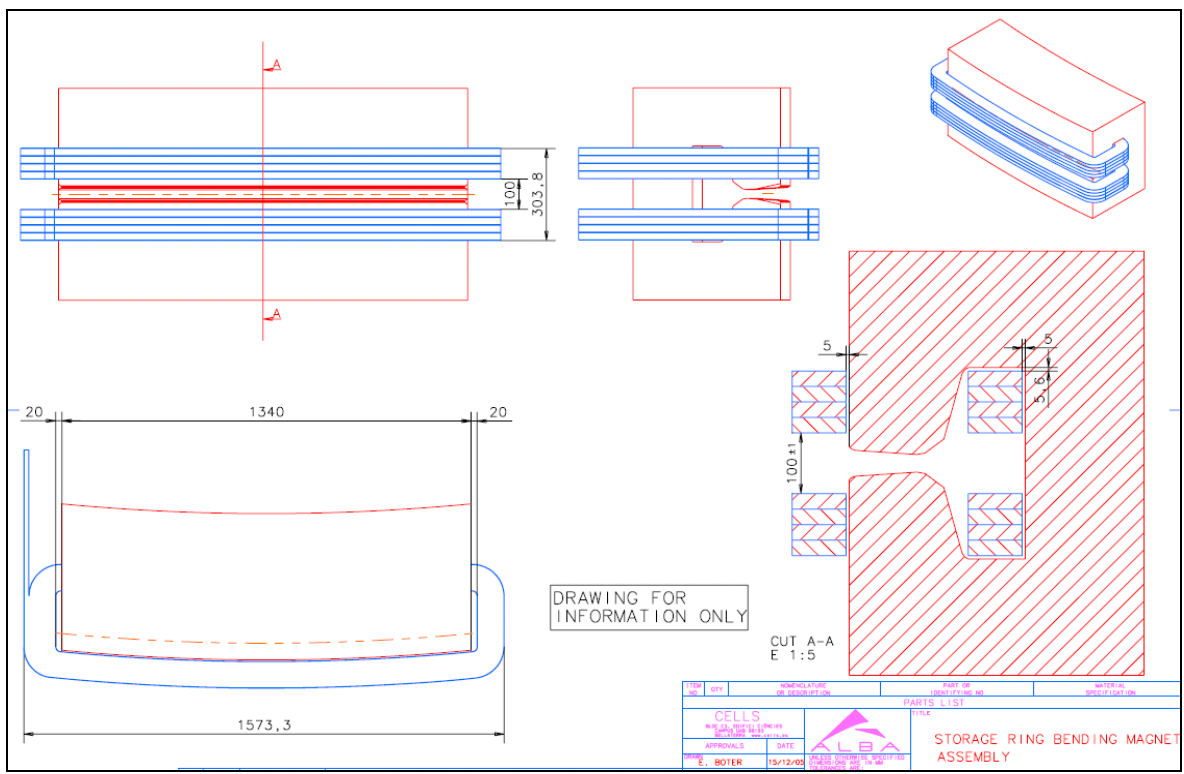

**Fig. 4:** Drawing showing the dimensions of the yoke and the coils for ALBA Storage Ring bending magnets

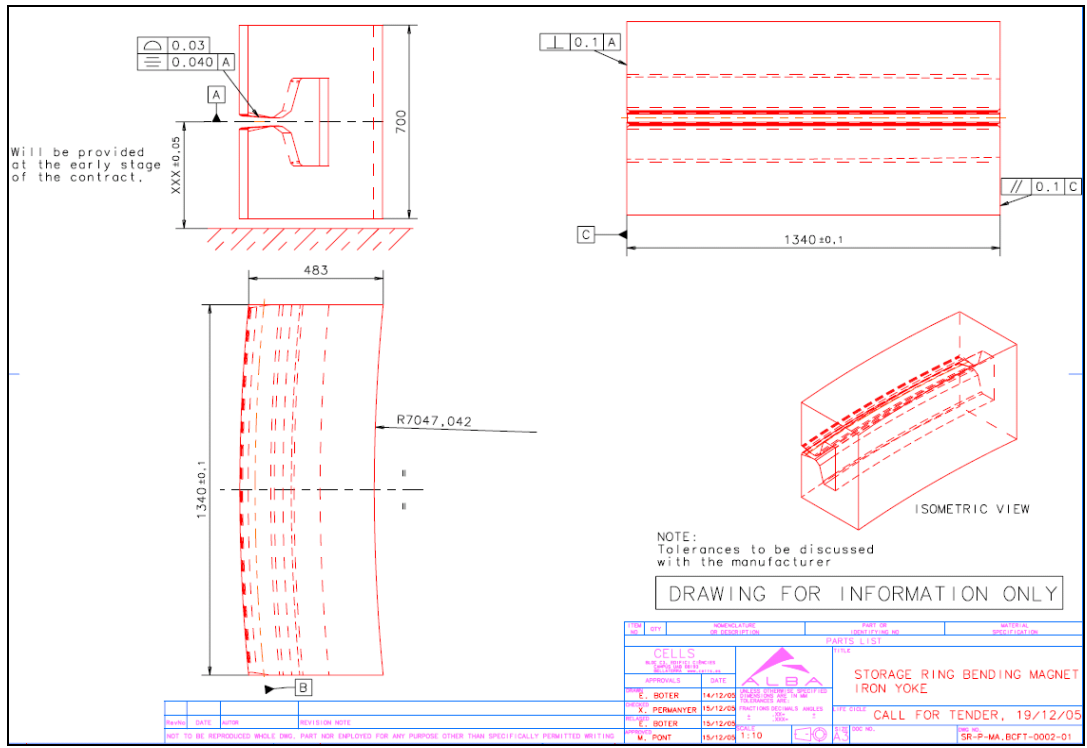

**Fig. 5:** Drawing showing the dimensions of the yoke for ALBA Storage Ring bending magnets

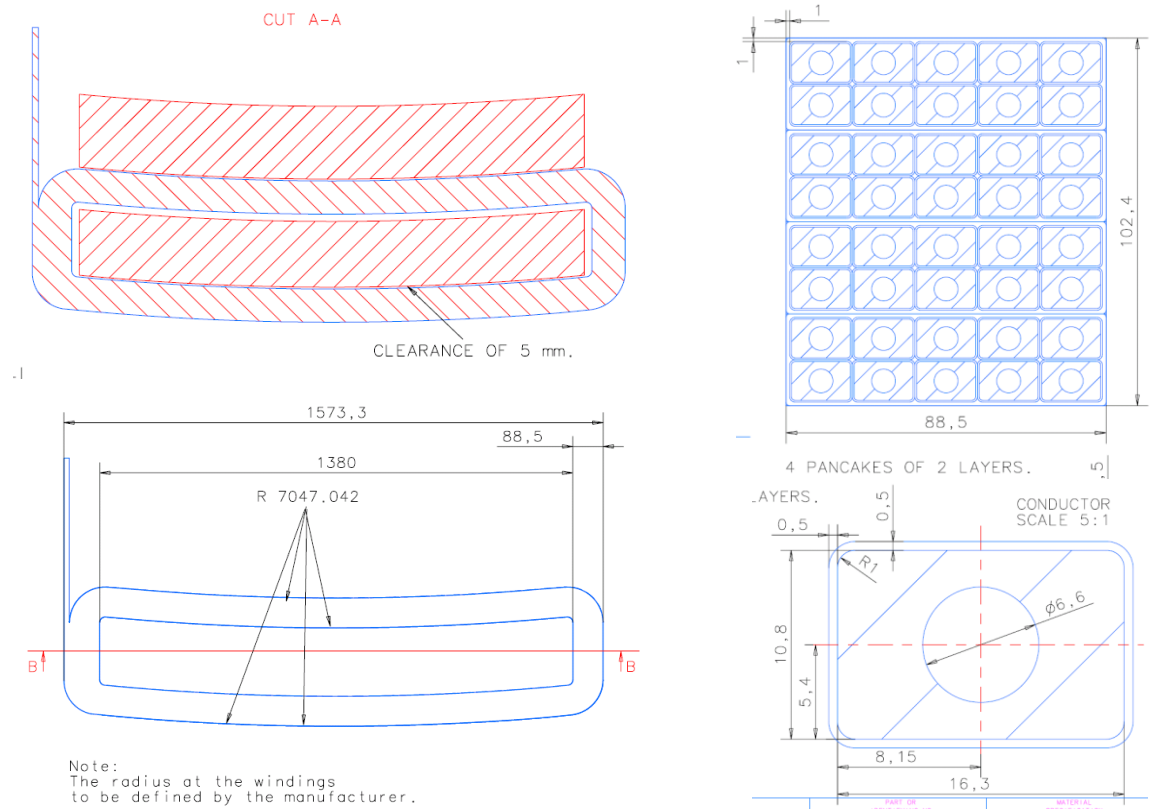

**Fig. 6:** Drawing showing the dimensions and characteristics of the coils for ALBA Storage Ring bending magnets

**Fig. 7:** Drawing showing the dimensions of end chamfers for ALBA Storage Ring bending magnet prototype. The end chamfer determines the overall length and the edge focusing of the magnet.

### 2.3.2.2 Quadrupole Magnets

Table 2 shows the list of specified parameters for the quadrupole magnets of ALBA Storage Ring and Booster accelerators. Figures 8–12 show some of the design drawings for Storage Ring quadrupole magnets.

**Table 2:** Parameter list of ALBA Storage Ring and Booster quadrupole magnets

| Storage Ring Quadrupole Magents | | | | | | Booster Quadrupole Magents | | | | |
|---|---|---|---|---|---|---|---|---|---|---|
| | | Q200 | Q260 | Q280 | Q500 | | | QS180 | QS340 | QC340 |
| **Magnetic properties** | | | | | | **Magnetic properties** | | | | |
| Beam Energy (E) | GeV | 3 | 3 | 3 | 3 | Beam Energy (E) | GeV | 3 | 3 | 3 |
| Field gradient (Go) | T/m | 19.8 | 21 | 21.4 | 21.9 | Field gradient (Go) | T/m | 17.45 | 17.45 | 17.45 |
| Sextupole component (B'') | T/(m^2) | 0 | 0 | 0 | 0 | Sextupole component (B'') | T/(m^2) | 0 | 0 | 5 |
| Effective length (Lo) | m | 0.23 | 0.29 | 0.31 | 0.53 | Effective length (Lo) | m | 0.2 | 0.36 | 0.36 |
| **Mechanical properties** | | | | | | **Mechanical properties** | | | | |
| Aperture radius | mm | 30.5 | 30.5 | 30.5 | 30.5 | Aperture radius | mm | 18 | 18 | 18 |
| Length of Fe-yoke L(Fe) | m | 0.2 | 0.26 | 0.28 | 0.5 | Length of Fe-yoke L(Fe) | m | 0.18 | 0.34 | 0.34 |
| Maximum length of magnet | m | 0.298 | 0.358 | 0.378 | 0.598 | Maximum length of magnet | m | 0.28 | 0.44 | 0.44 |
| **Coil and conductor** | | | | | | **Coil and conductor** | | | | |
| Number of coils | | 4 | 4 | 4 | 4 | Number of coils | | 4 | 4 | 4 |
| Number of turns per coil | | 46 | 46 | 46 | 46 | Number of turns per coil | | 17 | 17 | 17 |
| Conductor size | mm^2 | 8*8 | 8*8 | 8*8 | 8*8 | Conductor size | mm^2 | 5*5 | 5*5 | 5*5 |
| Cooling channel diameter (D) | mm | 5 | 5 | 5 | 5 | Cooling channel diameter (D) | mm | 3 | 3 | 3 |
| Number of ampere turns per coil | A-turns | 7327 | 7771 | 7919 | 8104 | Number of ampere turns per coil | A-turns | 2250 | 2250 | 2250 |
| Current (I) | A | 160 | 169 | 172.1 | 176.2 | Current (I) | A | 132.4 | 132.4 | 132.4 |
| Current density (j) | A/mm^2 | 3.59 | 3.81 | 3.88 | 3.97 | Current density (j) | A/mm^2 | 3.78 | 4.02 | 4.22 |
| Resistance at 23 degrees | mΩ | 51.9 | 60.8 | 63.8 | 96 | Resistance at 23 degrees | mΩ | 34.6 | 59 | 59 |
| Inductivity | mH | 24.6 | 32.4 | 34.6 | 59.2 | Inductivity | mH | 3 | 6 | 6 |
| Voltage drop | V | 8.26 | 10.3 | 11 | 16.9 | Voltage drop (resistive) | V | 4.6 | 7.8 | 7.8 |
| Power | kW | 1.32 | 1.73 | 1.89 | 2.98 | Power | W | 606 | 1034 | 1034 |
| **Cooling** | | | | | | **Cooling** | | | | |
| Maximim DT | Celsius | 8 | 8 | 8 | 8 | Maximim DT | Celsius | 8 | 8 | 8 |
| Nominal input temperature | Celsius | 23 | 23 | 23 | 23 | Nominal input temperature | Celsius | 23 | 23 | 23 |
| Number of cooling circuits per coil | | 2 | 2 | 2 | 4 | Number of cooling circuits per coil | | 1 | 1 | 1 |
| Maximum pressure drop per magnet | bar | 7 | 7 | 7 | 7 | Maximum pressure drop per magne | bar | 7 | 7 | 7 |

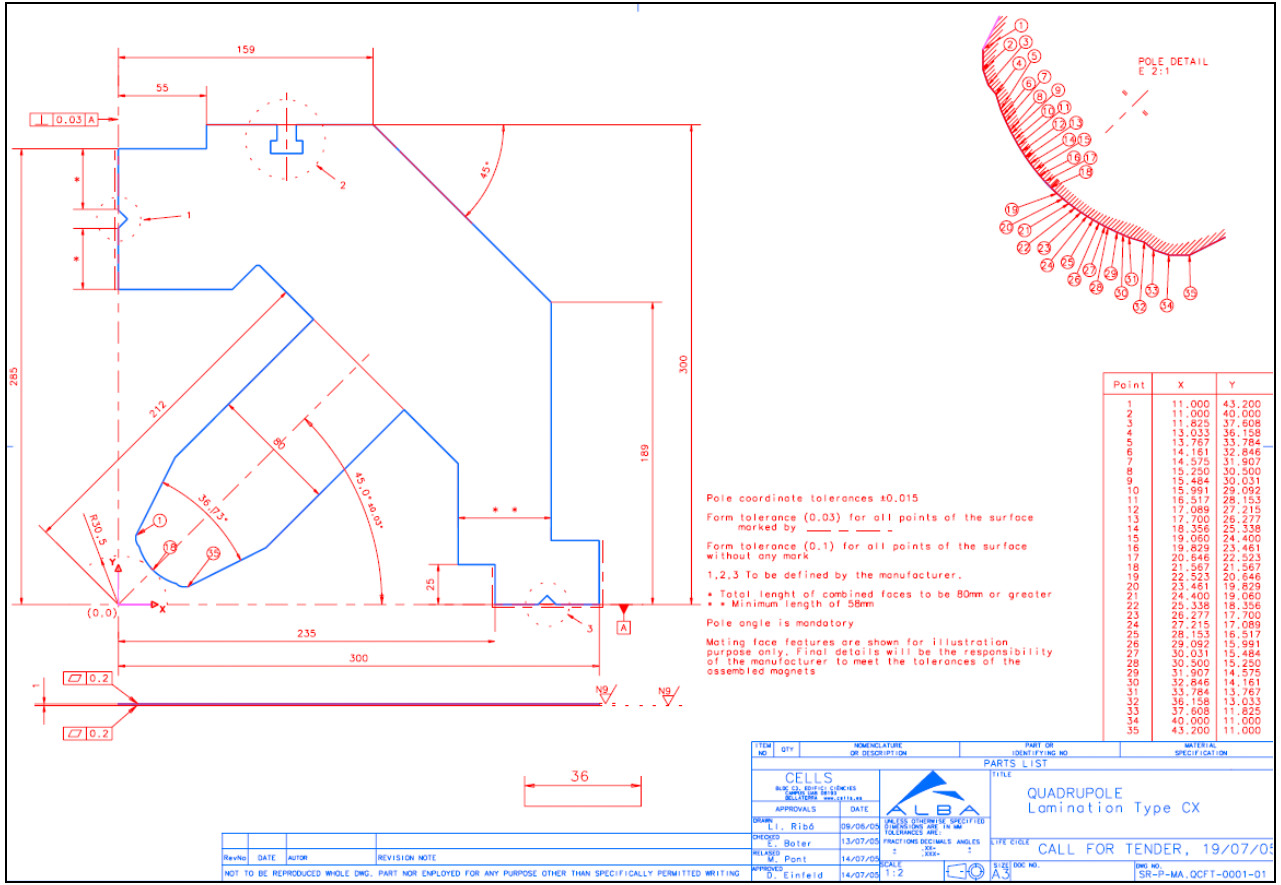

**Fig. 8:** Drawing showing the dimensions and the pole profile of the laminations for ALBA Storage Ring quadrupole magnets

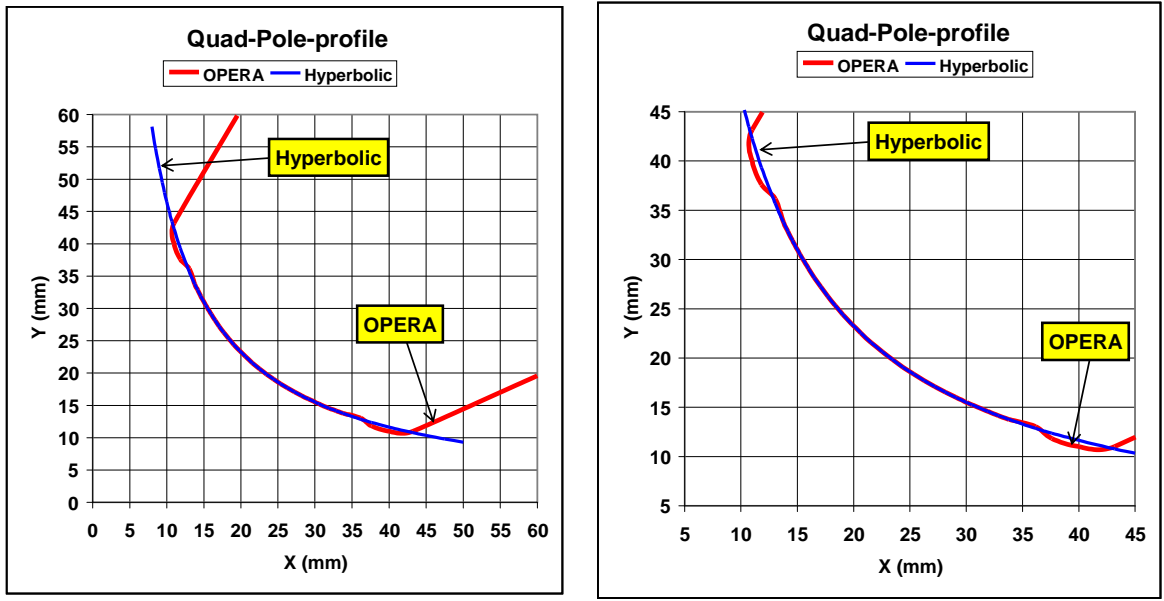

**Fig. 9:** Comparison between ideal (hyperbolic) pole profile and pole profile determined by means of OPERA magnetic field simulations corresponding to ALBA Storage Rings quadrupoles

**Fig. 10:** Dimensions and stay clear area for two types of ALBA Storage Ring quadrupole magnets

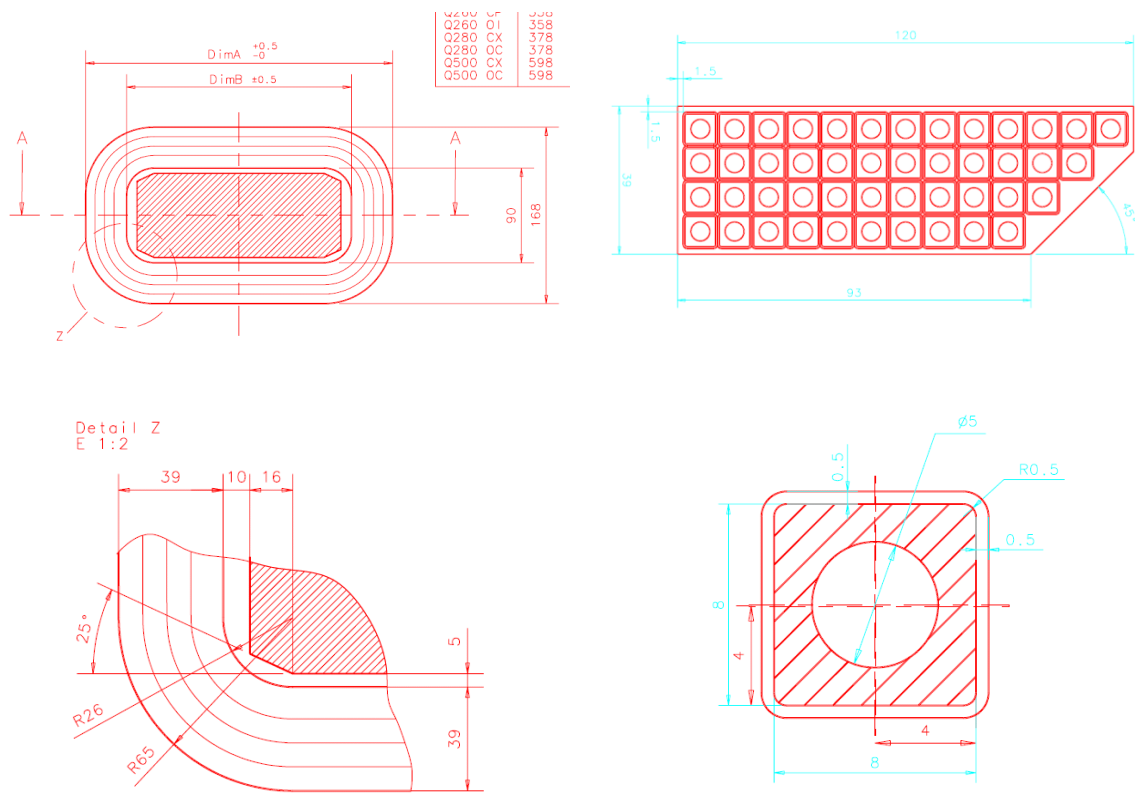

**Fig. 11:** Drawing showing the dimensions and characteristics of the coils for ALBA Storage Ring quadrupole magnets

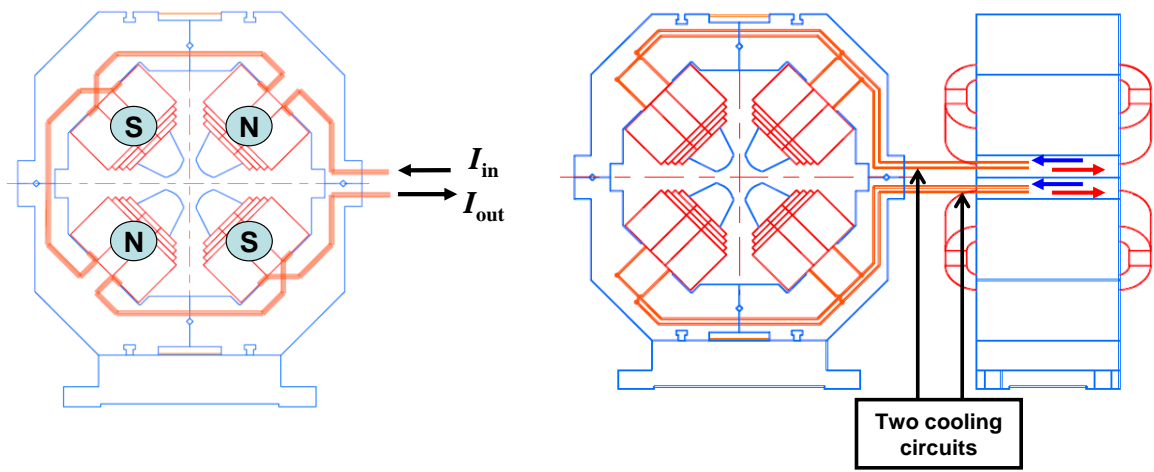

**Fig. 12:** Electrical (left) and water coolind (right) connections arrangement for ALBA Storage Ring quadrupoles

*2.3.2.3  Sextupole magnets*

Table 3 shows the list of specified parameters for the sextupole magnets of ALBA Storage Ring and Booster accelerators. Figures 13–15 show some of the design drawings for Storage Ring sextupole magnets.

**Table 3:** Parameter list of ALBA Storage Ring and Booster sextupole magnets

| Storage Ring Sextupole Magents | | | | Booster Sextupole Magents | | |
|---|---|---|---|---|---|---|
| | | S-150 | S-220 | | | S-200 |
| **Magnetic properties** | | | | **Magnetic properties** | | |
| Beam Energy (E) | GeV | 3 | 3 | Beam Energy (E) | GeV | 3 |
| Sextupole component (B'') | T/(m^2) | 700 | 700 | Sextupole component (B'') | T/(m^2) | 400 |
| Magnetic field at pole tip | T | 0.51 | 0.51 | Magnetic field at pole tip | T | 0.065 |
| Effective length (Lo) | m | 0.175 | 0.245 | Effective length (Lo) | m | 0.2 |
| | | | | | | |
| **Mechanical properties** | | | | **Mechanical properties** | | |
| Aperture radius | mm | 38 | 38 | Aperture radius | mm | 18 |
| Length of Fe-yoke L(Fe) | m | 0.15 | 0.22 | Length of Fe-yoke L(Fe) | m | 0.2 |
| Maximum length of magnet | m | 0.252 | 0.322 | Maximum length of magnet | m | 0.3 |
| | | | | | | |
| **Coil and conductor** | | | | **Coil and conductor** | | |
| Number of coils | | 6 | 6 | Number of coils | | 6 |
| Number of turns per coil | | 28 | 28 | Number of turns per coil | | 50 |
| Concductor size | mm^2 | 7*7 | 7*7 | Concductor size | mm^2 | 2.8*1 |
| Cooling channel diameter (D) | mm | 3.5 | 3.5 | Cooling channel diameter (D) | mm | |
| Number of ampere turns per coil | A-turns | 5094 | 5094 | Number of ampere turns per coil | A-turns | 310 |
| Current (I) | A | 182 | 182 | Current (I) | A | 6.2 |
| Current density (j) | A/mm^2 | 4.62 | 4.62 | Current density (j) | A/mm^2 | 2.21 |
| Resistance at 23 degrees | mΩ | 54.2 | 60.1 | Resistance at 23 degrees | mΩ | 886 |
| Inductivity | mH | 13.1 | 18.3 | Inductivity | mH | 34 |
| Voltage drop | V | 9.87 | 11 | Voltage drop (resistive) | V | 5.5 |
| Power | kW | 1.8 | 2 | Power | W | 33 |
| | | | | | | |
| **Cooling** | | | | **Cooling** | | |
| Maximim DT | Celsius | 9 | 9 | Maximim DT | Celsius | |
| Nominal input temperature | Celsius | 23 | 23 | Nominal input temperature | Celsius | |
| Number of cooling circuits per coil | | 3 | 3 | Number of cooling circuits per coil | | |
| Maximum pressure drop per magnet | bar | 5.6 | 7.46 | Maximum pressure drop per magne | bar | |

**Fig. 13:** Drawing showing the dimensions and the pole profile of the laminations for ALBA Storage Ring sextupole magnets

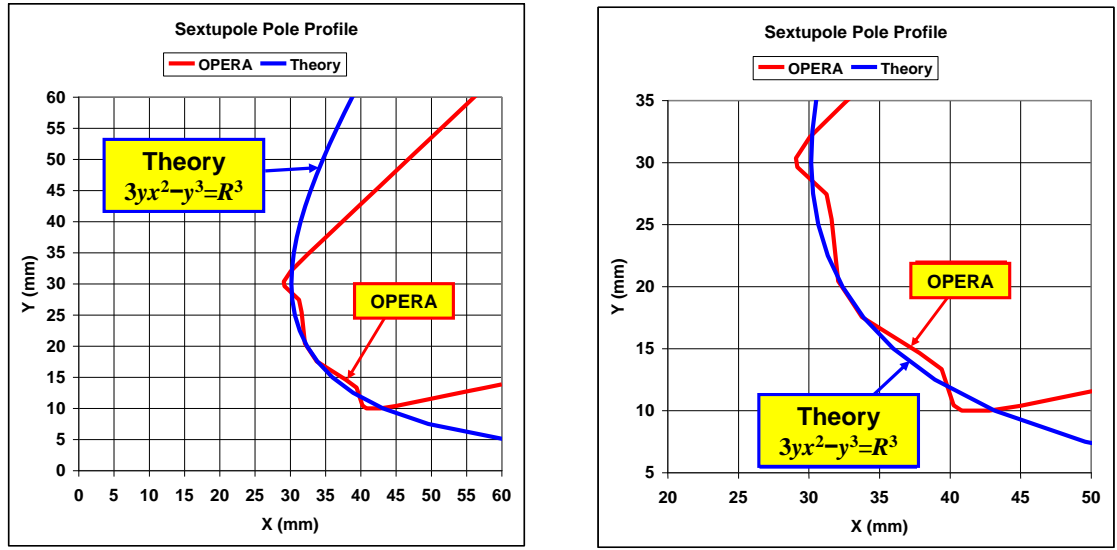

**Fig. 14:** Comparison between ideal pole profile and pole profile determined by means of OPERA magnetic field simulations corresponding to ALBA Storage Ring sextupoles

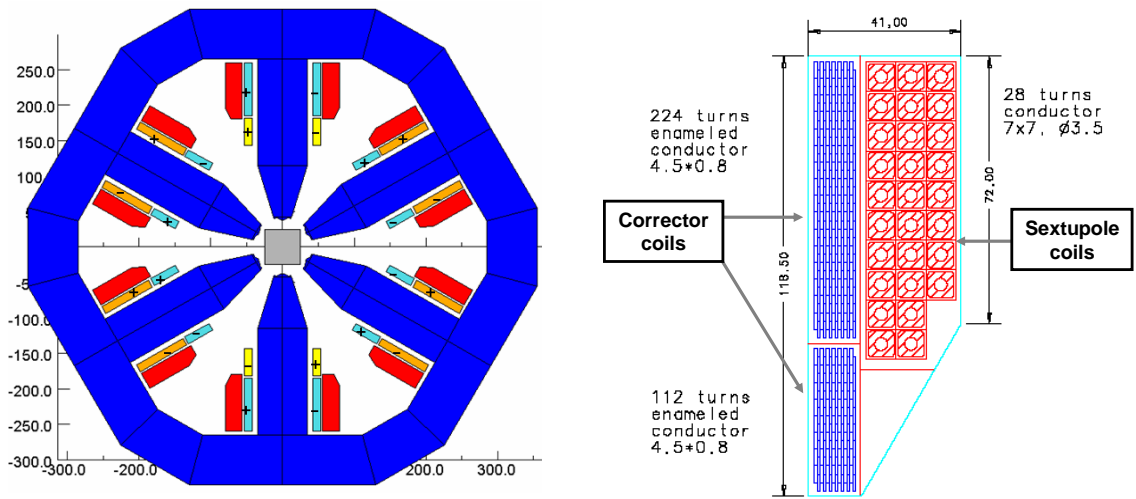

**Fig. 15:** Drawing showing the characteristics of the coils for ALBA Storage Ring sextupole magnets, which include additional windings for introduction of corrections (horizontal/vertical steering and skew quadrupole correction)

### 2.3.3  *Specification drawings*

2.3.3.1.) The drawings are specified in section 2.3.2. To produce a comprehensive set of drawings for the specification, it has been necessary to make certain assumptions concerning parts of the magnet design that will be the responsibility of the manufacturer. Tenderers shall therefore be aware that certain features shown on the drawings are tentative, and will be subject to adjustment by the manufacturer during the design phase. This reservation applies particularly to: (a) coil shape and cross section; (b) shape of magnet lamination, except for the pole profile.

### *2.3.4 Trim coils*

2.3.4.1.) The bending magnets shall be equipped with trim coils capable of a 5 per mil variation in the magnetic field (see Fig. 16). These coils are not shown in the drawings.

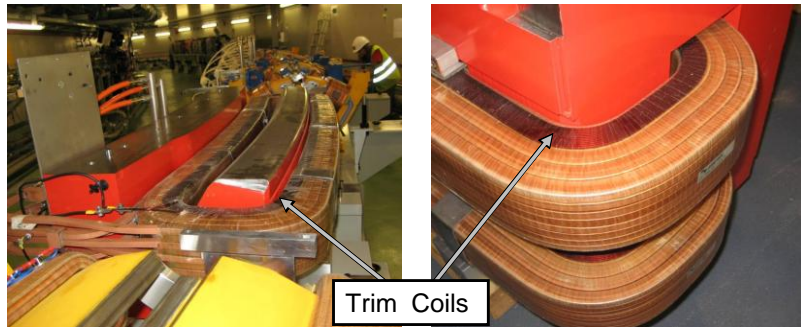

**Fig. 16:** Images of the trim coils of ALBA Storage Ring bending magnets

### *2.3.5 Magnet support feet*

2.3.5.1.) The bending magnets shall be supplied with supporting feet (see Fig. 17). Details of these feet, as well as of their location will be agreed with the manufacturer at an early stage of the contract.

### *2.3.6 Survey monument*

2.3.6.1.) The bending magnets shall be supplied with target mounting features on their upper face which are required to mount up to 3 survey monuments (see Fig. 17). Details of these monuments will be provided by CELLS at an early stage of the contract.

2.3.6.2.) The bending magnets shall be supplied with reference surfaces in the front and the side of the magnet (see Fig. 17). These surfaces shall be adequate for aligning the magnet with the magnetic bench by using, for example, a dial indicator. Details of these surfaces will be provided by CELLS at an early stage of the contract.

### *2.3.7 Lifting brackets*

2.3.7.1.) The bending magnets shall be supplied with at least four lifting brackets adequate to support the complete magnet (see Fig. 17).

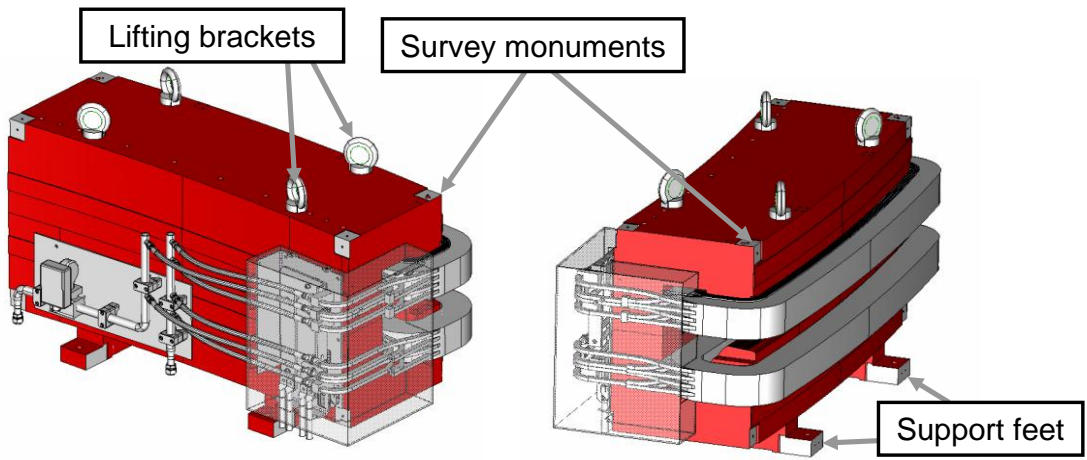

**Fig. 17:** Location of magnet support feet, survey monuments and lifting brackets for ALBA Storage Ring bending magnets

### 2.4 PHASING OF THE CONTRACT

2.4.1.) The contract will have two phases.

2.4.2.) The first phase includes:

- All mechanical and electrical design.
- The provision of all tools, jigs and fixtures.
- The procurement of all materials for one (1) pre-series magnet.
- The procurement of all materials that are not available "ex-stock" for the series production magnets.
- The manufacture, mechanical and electrical testing of one (1) pre-series magnet.
- The manufacture of 3×2 different sets of end chamfers as shown on the appropriate drawing.
- The mechanical measurement and testing of the pre-series magnet, as specified in section 2.10.1.

2.4.3.) CELLS will perform the magnetic measurements of the pre-series magnet with the provided end chamfer pieces in order to define the final end chamfer to be used on the production series magnets and asses the viability of the magnet design.

2.4.4.) The second phase includes:

- Any necessary modification of the tooling, which results from the mechanical and magnetic testing of the pre-series.
- The provision of all additional materials required for the production of magnets.
- Manufacture, electrical and mechanical testing and delivery of thirty-two (32) bending magnets.
- Manufacture, testing and delivery four (4) sets of spare coils.

2.4.5.) CELLS will not authorise the second phase until tests and measurements on the pre-series magnet have been successfully completed by the manufacturer and approved by CELLS.

2.4.6.) All designs, tools and materials obtained during phase one shall remain the property of CELLS, and shall be surrendered to CELLS at any time during the contract, within one month of the receipt of written notification.

2.4.7.) Tenderers are required to provide separate quotations for the two phases, which together will fully cover the 'scope of contract' defined above.

2.4.8.) The quotation for both phases of the contracts will be fixed price.

## 2.5 SCHEDULE

2.5.1.) The following program is required for the timescales of the design, construction and delivery:

| | |
|---|---|
| Contract award | Week 0 |
| Engineering design ready (including tooling design) | Week 16 |
| Pre-series magnet completed | Week 32 |
| Completion of pre-series tests at the factory | Week 34 |
| Magnetic measurements at CELLS and approval of pre-series | Week 42 |
| Completion of delivery of production bendings | Week 66 |

## 2.6 TENDERING AND CONTRACT MANAGEMENT

### 2.6.1 Tendering

2.6.1.1.) All interested contractors are strongly encouraged to contact CELLS and discuss details of the specification to ensure that the bidder understands completely the requirements and implications of the specification before making an offer. Enquiries of a technical nature shall be directed to X.YYYY, CELLS, tel: xx-xx-xxxyyyy, e-mail: x.yyy@cells.es. Enquiries of a contractual nature shall be directed to Mr. z.wwwwww, CELLS, Tel: xx-xx-xxxwwww, e-mail: z.wwwwww@cells.es.

2.6.1.2.) CELLS shall adjudicate the bids by considering the technical, and value for money aspects of the formal bid. See the folder of administrative clauses.

### 2.6.2 Information required with the tender

2.6.2.1.) The bidder shall provide with the tender documents sufficient information to allow an informed choice of contractor. These shall include:

- A confirmation of acceptance of every clause of the present specification or a detailed explanation of any departure from the conditions defined in this specification.

- A breakdown of the price into main categories.

- Details of the quality assurance scheme that the contractor operates.

- A draft time schedule showing the principal design, ordering, manufacturing, and testing of the bending magnets.

- Indications of proposed work packages to be undertaken by any sub-contractors with the identity of the proposed subcontractor.

- A list of previous projects, similar or comparable in size and scope, to enable CELLS to assess the contractors viability and ability to accomplish the contract.

2.6.2.2.) Specific information on:

- Engineering Design
  – Where major features of the magnet design have been left to the judgement of tenderers, an indication of the solutions or parameters shall be given.
  – Proposed method to hold the yoke together.
  – Proposed method to keep the strict mechanical tolerances on the gap region as set in the corresponding drawings.

- Magnetic Steel
  - Proposed source of supply of the steel that will be used in the magnet yokes
  - Technical information required for the proposed steel shall include: Thickness / Grade or type designation / Proposed nature of insulated coating / Quoted permeability / Quoted coercivity.
  - The tenderer shall indicate the test and measurement methods that are proposed for quality control of the mechanical, electrical and magnetic properties of the magnet steel.
- Lamination Stamping and Yoke Assembly
  - A brief description of the tooling proposed for stamping laminations, and whether a one or two stage process is intended shall be given. The size of the stamping press to be used shall be indicated.
  - An estimate of the burr height and extent of shear edge taper expected on the lamination, together with the estimated frequency of tool regrinding to meet the specification, shall be given.
  - Details of the methods proposed and the equipment available for the accurate measurement of the lamination profile.
  - Details of the storage procedure and the method of the subsequent shuffling operation.
  - A description of proposed lamination stacking fixture, including stacking technique. The proposed method for holding the laminations together, and the techniques recommended for maintaining the required dimensional tolerances on the yoke shall be explained.
- Lamination Stamping and Yoke Assembly
  - Details of the measurement procedures proposed for mechanical checking of the yoke after stacking and bonding/welding, and indication of how the dimensional tolerances given by CELLS can be checked.
  - Details of the proposed techniques for any required machining.
  - A preliminary proposal of the instrument that is to be used for the continuous measurement of the gap, taking into account that the gap is not constant in the transversal direction.
- Coil Production and Testing
  - Indication of the expected source of supply for copper conductor, and details of the proposed conductor dimensions.
  - Details of the proposed coil winding operation, including information on the glass cloth to be used. The method of production and all materials to be used (including required packing pieces) must be described. The number of coil winding formers to be used must be stated.
  - The tenderer shall indicate the proposed source of supply of the epoxy resin chemical system, together with details of its expected mechanical, thermal and radiation properties.

- Details of the proposed impregnation technique and curing operation for the resin system must be given, including information on all jigs, tools and moulds which will be required. The number of moulds to be used must be stated.
- A brief description shall be given of the equipment, which is proposed for the various tests listed in the coil test schedule.
- Details of the proposed over-temperature switches shall be given. The proposed method of mounting the temperature sensor on the coil shall also be explained.
- Details of the proposed water flow switch shall be given. The proposed method of mounting the sensor on the magnet shall also be explained.

- Electrical Connections
  - The tenderer shall indicate which type of connections are proposed for the power cables as well as the proposed layout of terminal boards and protective covers.
  - The tenderer shall indicate the proposed method of coil electrical interconnection and the way that radiation damage is to be avoided.
  - The tenderer shall give details of the proposed connections of the interlock switches that are proposed.
- Water Connections
  - The tenderer shall give details of the proposed water distribution, including drawings of the manifolds and indicating which materials will be used.
  - The water flow rate shall be given.
- The proposed layout, avoiding organic materials in the bending magnet vertical median plane shall be described.

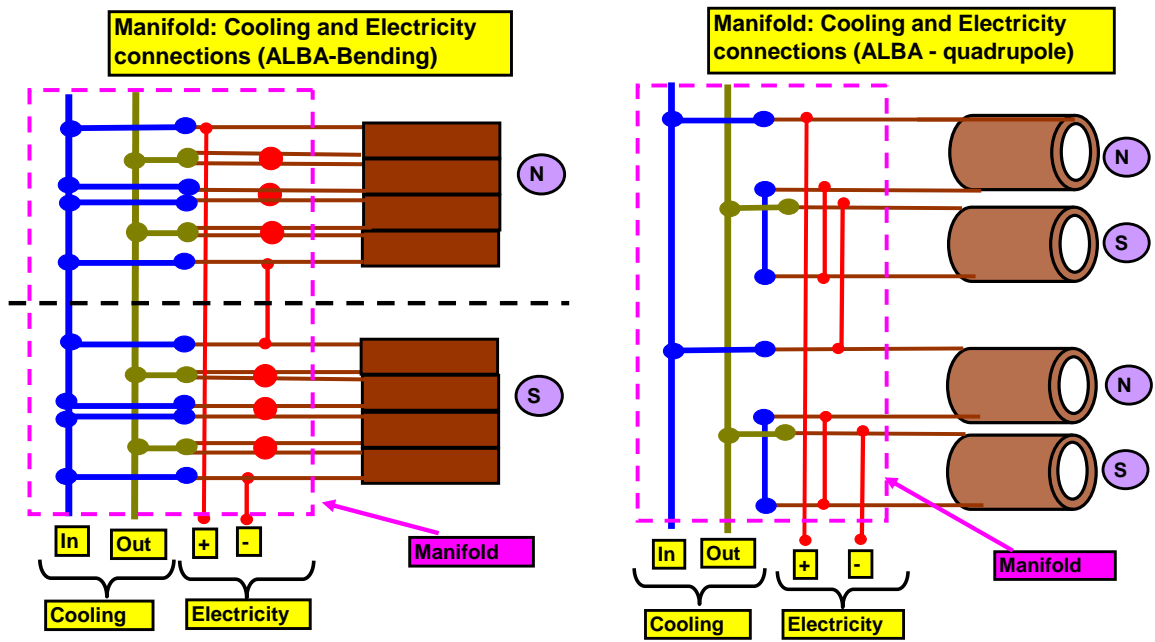

**Fig. 18:** Scheme of water and electrical connections for ALBA bending and quadrupole magnets

### *2.6.3 Contract management*

2.6.3.1.) At the start of the contract the contractor shall assign contact persons for Technical and Administrative matters who will be responsible for all reporting to, and contact with CELLS.

2.6.3.2.) Within 2 weeks of the commencement of the contract the contractor must issue a detailed program covering the design, manufacturing, and testing phases in sufficient detail to allow regular progress monitoring.

2.6.3.3.) Within 2 weeks of the commencement of the contract a program of technical and progress meetings will be agreed between the contractor and CELLS.

2.6.3.4.) Thereafter, and throughout the contract, the Technical Contact shall supply a written report to CELLS every month detailing progress with respect to the program.

2.6.3.5.) The manufacturer will be responsible for the final design, the production methods and the correct manufacture of all magnets, irrespective of whether they have been chosen by the contractor or suggested by CELLS.

2.6.3.6.) CELLS's approval of the design and components does not release the contractor from his responsibility to correct errors, oversights and omissions to ensure conformance to the specifications in this document.

2.6.3.7.) In the event of the contractor having misinterpreted any of the specifications provided by CELLS, CELLS expects that the misinterpretation will be corrected at no extra cost.

2.6.3.8.) The contractor must declare any sub-contractor that will be used in the execution of this contract and inform CELLS of any change of sub-contractor. The change must be accepted by CELLS in writing.

2.6.3.9.) Nominated members of CELLS staff, or their appointed representatives must be guaranteed reasonable access to the premises of the contractor, for the purpose of progress meetings, inspection visits etc., even without notice.

2.6.3.10.) Nominated members of CELLS staff, or their appointed representatives must be guaranteed reasonable access to the premises of any subcontractor, for the purpose of progress meetings, inspection visits etc., with the main contractor present, even at short notice.

2.6.3.11.) CELLS will have the right to observe the tests and suggest additional ones. The contractor shall give at least 10 working days notice of any test date to allow the necessary travel arrangements to be made.

2.6.3.12.) CELLS reserves the right to require additional or more extensive tests to be conducted in the event of marginal design or performance.

2.6.3.13.) The contract will be completed when the magnets have been delivered to the CELLS indicated site and satisfactorily completed the acceptance tests and comply fully with this specification document.

### *2.6.4 Quality assurance*

2.6.4.1.) The contractor shall maintain and apply a quality assurance program compliant with ISO-9001 for the design, manufacture and testing of all systems and equipment provided by them.

2.6.4.2.) All equipment shall be manufactured in accordance with the best existing techniques and recognised good engineering practices available at the time of construction. All systems shall be designed and constructed with an expected operational lifetime longer than 10 years. The magnets shall be designed and constructed for continuous use.

### 2.6.5 *Documentation*

2.6.5.1.) The contractor shall provide 2 (two) sets of paper copies of the following documentation as soon as it becomes available and in accordance with the schedule presented in section 5:

- Schedule
- Construction drawings of coils and magnetic circuit
- Travellers of control executed at each step of the manufacturing
- A manufacturing booklet which will be composed of:
  - For the coils: Mechanical, hydraulic and electric tests
  - For the magnetic yoke: Dimensional checks
  - For the magnet assembled: Mechanical, hydraulic and electric tests
  - The list of the non compliances processed

2.6.5.2.) All the technical documents delivered by the contractor shall be in English.

2.6.5.3.) The contractor shall provide 2 (two) full sets of paper copies of construction drawings.

2.6.5.4.) In addition, the contractor shall provide 2 (two) full sets of electronic copies on physical media of construction drawings. These shall be preferably in IDEAS format, although other formats e.g. DXF, or IGES are acceptable.

### 2.6.6 *Numbering*

2.6.6.1.) Each individual coil, yoke and each completed magnet will be identified and numbered. The position of the identification number shall be agreed with CELLS.

2.6.6.2.) A stainless steel or aluminium plate shall be fixed on each magnet. The following information shall be on the plate:

- Magnet name & Serial number
- Year of manufacture & Maximum current
- Gross weight of the unit & Cooling requirements (if required)
- Coolant flow rate (l/min) & Pressure drop (MPa)
- Maximum Temperature Rise (K) & Contractor's name

### 2.6.7 *Guarantee*

The contractor shall guarantee the magnets against failure due to either faulty components or faulty manufacture for a period of 24 months after the magnets have been accepted by CELLS. It is warranted that no modifications will be undertaken without the written permission of the contractor.

## 2.7 TECHNICAL SPECIFICATIONS: MAGNETIC STEEL

### 2.7.1 *Steel characteristics for the magnet*

2.7.1.1.) It is envisaged that this specification will be met by cold rolled, fully annealed, non-oriented, laminated steel.

2.7.1.2.) Table 4 below gives the minimum values of induction under d.c. excitation acceptable at the stated values of field parallel to the rolling direction; values of relative permeability are shown for convenience. Tests are assumed to be made on strips, so that properties parallel and perpendicular to the rolling direction can be separately assessed. Induction measured perpendicular to the rolling direction shall be not less than 20% lower than the values given in Table 4.

Table 4: Required induction and permeability for the magnet steel

| Magnetic field [A/m] | Minimum induction parallel to rolling direction [T] | Relative d.c. permeability |
|---|---|---|
| 116 | 0.50 | 3430 |
| 208 | 1.00 | 3826 |
| 300 | 1.30 | 3448 |
| 597 | 1.50 | 1999 |
| 1343 | 1.60 | 948 |
| 3236 | 1.70 | 418 |
| 6855 | 1.81 | 210 |
| 12490 | 1.91 | 122 |

2.7.1.3.) The coercitivity is defined as the field required to produce zero induction in a mixed sample of the steel after repeated cyclical excursions to high induction with a field of $\geq 10\,000.0$ A/m. The maximum allowed coercitivity in a single sample is 80 A/m, and the maximum allowed variation from the mean is ±15 %.

2.7.1.4.) The steel magnetic properties shall be guaranteed on bulk samples after stamping without any further heat treatment or annealing.

2.7.1.5.) The steel is required to be coated, at least, on one side with an inorganic insulating coating with a maximum thickness of 5 μm.

## 2.7.2 Laminations

2.7.2.1.) Nominal thickness of the lamination is 1 mm.

## 2.7.3 Testing of steel

2.7.3.1.) Steel is normally produced in 'batches'. All batches of steel produced by the steel suppliers shall have test samples taken from the beginning and end of the batch, together with a further sample from the middle of the batch. In exceptional circumstances, where tests on these samples indicate that a large variation of magnetic properties, greater than 15% peak-to-peak, is present within a single batch, CELLS shall be entitled to call for further samples to be taken at one quarter and three quarters through the batch.

2.7.3.2.) All three samples of each batch shall have the following measurements carried out with a strip sample technique for properties parallel and normal to the rolling direction

    i) Permeability at all values of induction specified in section 2.7.1.2 (see Table 4).

    ii) Coercivity as defined in section 2.7.1.3.

2.7.3.3.) The surface insulation shall be checked on three samples taken at locations indicated above, using the standard insulation measuring technique of the steel manufacturer.

2.7.3.4.) The thickness of the steel shall be checked on samples, taken at locations indicated above.

2.7.3.5.) The tenderers shall provide information at tender on the proposed methods for magnetic, electrical and physical measurements of the magnet steel.

### *2.7.4 Steel supplier*

2.7.4.1.) Steel of the described quality is available from a number of suppliers and the manufacturer has full liberty to choose any source of suitable material. CELLS believes that e.g. steel type 1200-100A coated on both sides with Stabolit 70 will comply with this specification. CELLS has identified the supplier for this steel:

*ThyssenKrupp Electrical Steel GmbH (EBG), Altendorfer Str. 120,D-45143 Essen, Germany*

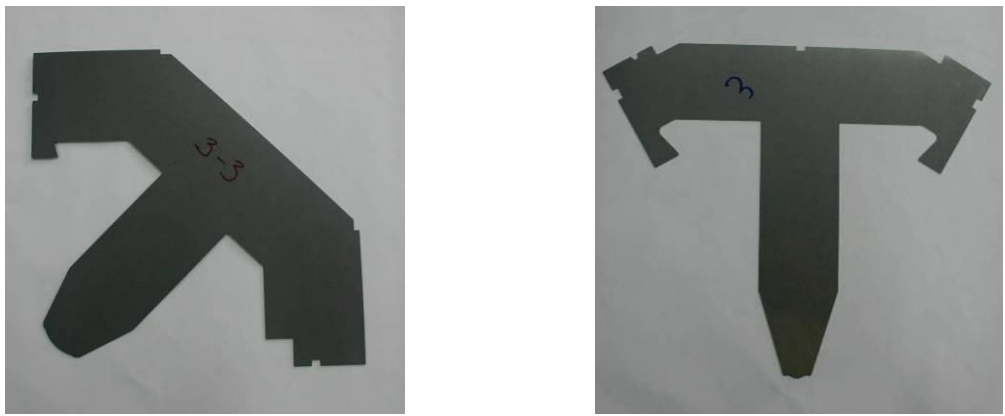

**Fig. 19:** Examples of laminations for ALBA Storage Ring quadrupoles (*left*) and sextupoles (*right*)

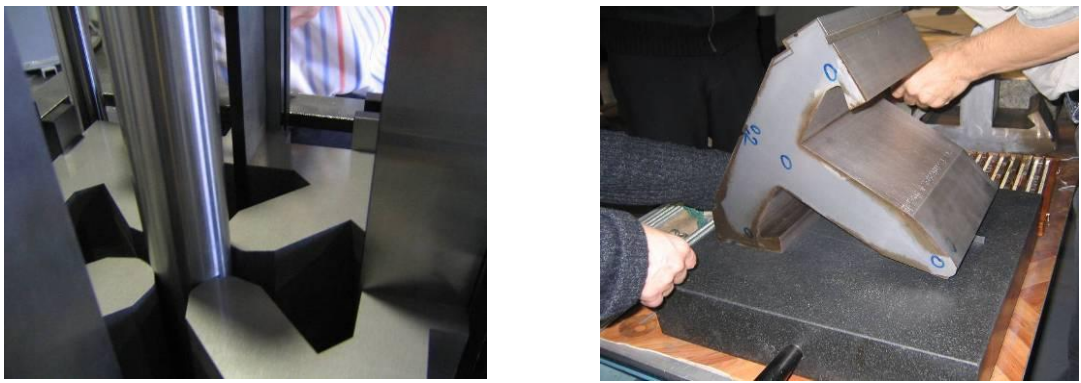

**Fig. 20:** *Left:* Stacking fixture of laminations for ALBA Storage Ring quadrupole. *Right:* Mechanical test of finished yoke for ALBA Storage Ring sextupole magnet.

## 2.8 LAMINATIONS AND YOKES

### *2.8.1 Laminations stamping*

2.8.1.1.) The manufacturer is responsible for achieving the required dimensional tolerances for all laminations to meet the mechanical specifications.

*2.8.2  Lamination stamping tests*

2.8.2.1.) Prior to start production stamping and after each re-grinding of the stamping tools, a certain number of lamination will be punched and three of them shall be measured. In addition, after every 5000 laminations during the stamping operation, a lamination will be taken at random from the production line, and measured. The suitability of the lamination will be judged with respect to the lamination drawing that will have been approved by CELLS. The measurements shall be submitted to CELLS for approval.

2.8.2.2.) CELLS reserves the right to be present during all such measurements.

2.8.2.3.) The tested laminations, duly marked with an identifying label, shall then be made available to CELLS for independent verification of dimensional tolerances.

2.8.2.4.) CELLS is entitled to reject any lamination which is not within tolerances, as specified in the appropriate drawing.

2.8.2.5.) The maximum burr at any location on the lamination must not exceed 0.025 mm; no subsequent deburring operation will be permitted.

2.8.2.6.) Re-grinding of the tooling must be carried out at such intervals as are necessary to maintain the required dimensional tolerances of all laminations.

2.8.2.7.) After stamping, the laminations produced from each batch of steel shall be stored separately to allow the shuffling operation as described below to be undertaken. The batch of origin of all laminations shall continue to be identifiable.

2.8.2.8.) In order to successfully carry out the shuffling operation, all steel delivered for use in Phase 2 of the contracts must be stamped and stored before commencing any further handling and processing operations on the laminations. The steel shall be protected against corrosion.

*2.8.3  Laminations shuffling*

2.8.3.1.) This section applies only to laminations produced for use in the second phase.

2.8.3.2.) The required magnetic identity between the magnets requires appropriate shuffling of the laminations in such a way that each magnet contains steel representative of each steel production batch, in roughly equal proportions. Manufacturers shall therefore plan a strategy to achieve this, taking into account the number of steel batches required for the magnet production, and the expected variation of magnetic properties within each batch.

2.8.3.3.) Final details of the shuffling process shall be agreed with CELLS; these will depend on the results obtained during the magnetic testing of the steel and the periodic measurement of the laminations defined in section 2.8.2.1.

*2.8.4  Yoke*

2.8.4.1.) Tenderers shall indicate whether it is their intention to stack and glue laminations to produce yokes in a final state or whether it is their intention to stack and weld the laminations to produce a yoke.

2.8.4.2.) Tenderers shall indicate whether it is their intention to perform further machining on one or more faces of the block after yoke assembly. Where such machining is proposed, tenderers shall give a clear explanation of how they intend to ensure that the specified dimensions required on the yoke will be achieved. In any case machining of any part of the yoke is restricted to:

  a) Machining of the pole profile

b) Machining of the reference planes for the alignment features, as set out in section 2.3.6.

c) Machining of the reference for the support feet.

The pole has one machined end chamfer to correct the end field in order to provide the correct harmonic analysis. After magnetic measurements of the pre-series magnet the final shape of the end chamfer will be confirmed.

2.8.4.3.) During the stacking of the initial pre-series, a record shall be kept of the weight of laminations used for the yoke. After acceptance of the pre-series, this weight will be defined as the standard weight of laminations in a given yoke.

2.8.4.4.) During stacking of the production magnet yokes, the number of laminations to be included in the yoke will be determined by weight, and shall be within plus or minus half the weight of a single lamination of the standard weight of laminations in a yoke block.

2.8.4.5.) The packing factor has to be at least 98 % and has to be within ±0.5 % for all of them.

### 2.8.5 Bending magnet ends

2.8.5.1.) To control both the effective length and the integrated field quality of the bending magnet, the ends of the poles of each magnet will be chamfered, as specified on the appropriate drawing.

2.8.5.2.) To decide the length and the angle of the end chamfer a set of different chamfers, as specified on the appropriate drawing, will be produced during the first phase and a set of magnetic measurements will be taken for the pre-series magnet equipped with all different chamfers. After evaluation of the magnetic measurements done by CELLS the length and angle of the end chamfer to be machined on the production magnets will be communicated to the manufacturer before the second phase starts.

2.8.5.3.) Irrespective of the technique used for producing the end modification, the laminations that are to be included in the end region of the poles will be taken from the same stock of shuffled laminations as the unmodified pole region.

### 2.8.6 Mechanical yoke testing

2.8.6.1.) After the laminations have been assembled and machined into a yoke the main geometrical dimensions will be carefully checked according to the tolerances of the relevant drawings. Features that shall be controlled are:

a) Length measured at three different locations along the yoke

b) Flatness of the whole assembly

c) Squareness of the sides and of the end faces

d) Longitudinal shape of the yoke will be checked with a jig, the construction of which is part of this contract.

2.8.6.2.) After assembling of the complete yoke the functional tolerances among which the distance between the reference surfaces on the top and bottom of the bending magnets and the median plane will be checked with the magnet both powered and unpowered. The dimensional controls are specified on the appropriate drawings.

### 2.8.7 Protection and painting

2.8.7.1.) After assembly and control, the yokes will be protected against rust by painting. Two-component epoxy paint shall be used, which the manufacturer shall ensure is hard and mechanically resistant. The unpainted areas, as indicated in the appropriate drawings shall be protected by a light oil or other rust preventative measures. The colour will be RAL 3026.

## 2.9 TECHNICAL SPECIFICATIONS: COILS

### 2.9.1 Coils manufacturing

2.9.1.1.) Bending magnets have two (2) coils. Each coil is made of four (4) individual pancakes. Each pancake is composed of 10 turns; therefore, each coil has 40 turns. The coils shall be manufactured using a solid copper conductor with a central hole for the passage of cooling water, according to the CELLS design.

2.9.1.2.) The coil is designed to have a maximum temperature rise in the cooling water of 10°C with a differential pressure of 7 bar. Under these conditions the maximum coil temperature will be ≤35°C. It is essential that the insulation system withstands repeated thermal cycling without mechanical or electrical failure.

2.9.1.3.) The coils will operate in a radiation environment, and must therefore be built using glass and epoxy resin insulation system. No materials other than those specified in this section of the specification will be permitted. The materials that will be used for insulation will be subject to written authorization from CELLS.

2.9.1.4.) Inter-turn resistance shall be provided by wrapping the copper conductor with a borosilicate glass cloth, half-lapped to produce a minimum insulation thickness of 0.5 mm turn-to-turn. After completion of winding, an outer ground insulation shall be provided by further layers of half-lapped glass cloth of minimum thickness 1.0 mm.

2.9.1.5.) On completion of the coil winding, electrical and water terminations shall be attached to the leads.

2.9.1.6.) Each pancake shall have one (normally closed) over-temperature switch, set to open an electrical circuit at 60˚±5˚C. These switches shall be fitted each with two external leads or connections, and shall either be fitted before the impregnation process, or glued to the surface of the coil after impregnation. Whichever technique is used, the manufacturer will ensure that good thermal and mechanical contact is obtained using materials that meet the requirements of this specification.

2.9.1.7.) All voids arising within the pancakes shall be packed with glass roving in order to avoid the occurrence of resin-rich areas and delamination. The material used shall conform to the glass requirements described below.

2.9.1.8.) The insulation in the vicinity of the pancakes leads and terminations will require special attention, in order to provide adequate strength and to avoid the presence of excessive resin in that area. Glass roving or pre-formed glass epoxy inserts must be utilised in these areas, and all materials used must conform to the requirements described below.

2.9.1.9.) The manufacturer shall estimate the degree of conductor keystoning that will occur in certain areas of the coil. Resulting voids consequently introduced into the coil must therefore be filled, using the methods indicated above in 2.9.1.8.

2.9.1.10.) No joints within the pancakes will be permitted.

### 2.9.2 Conductor

2.9.2.1.) The copper shall be Cu-OF Oxygen free (ISO designation) annealed after cold work (dead soft fully annealed temper).

2.9.2.2.) The copper shall be free of cracks, porosity and voids. It shall not have any tendency for hydrogen embrittlement. Very good characteristics for brazing are required as well as a ductibility which permits the winding of the conductor into magnet coils with tight bends.

2.9.2.3.) The composition shall be at minimum 99.99 % Cu (+Ag).The oxygen content shall be kept below 10 ppm.

2.9.2.4.) The electrical resistivity shall be less than $17.1 \times 10^{-9}$ $\Omega \cdot$m at 20°C.

2.9.2.5.) The uniformity of the conductor shall be such that the resistance of all coils constructed from it shall be equal to within ±1%.

### *2.9.3 Conductor supplier*

2.9.3.1.) Copper conductor of the described quality is available from a number of suppliers and the manufacturer has full liberty to choose any source of suitable material. CELLS has identified a supplier for a conductor that will comply with this specification:

*Luvata, Kuparitie, P.O.Box 60, FIN-28101 Pori, Finland, www.luvata.com*

### *2.9.4 Conductor tests before winding*

2.9.4.1.) Test certificates shall be available relating to tests undertaken by the copper manufacturer, to include dimensions, resistivity and Brinell hardness.

2.9.4.2.) The cooling channel must permit the free passage of a 5.3 mm diameter ball.

2.9.4.3.) Before construction commences, the conductor shall be hydraulically tested at a pressure of 100 bar for five minutes. Conductors revealing any evidence of leakage shall be rejected.

2.9.4.4.) Before the winding of the coil, the conductor shall be cleaned and sandblasted.

### *2.9.5 Pancake winding*

2.9.5.1.) Scrupulous care shall be exercised at all stages of the coil construction in the handling of all the components, which shall be undertaken in a clean environment. All working surfaces shall be cleaned immediately prior to be used, and protective gloves shall be worn by all the staff involved. Quality control of this design will have to guarantee the lack of any conductive occlusion between wires (cutting, dust…). Moreover, excessive hammering (hardening) of the conductor which could destroy the fibreglass tape, shall be avoided.

2.9.5.2.) After completion of pancake winding, the pancakes shall be tested with a gas (helium or halogen) at a pressure of 15 bar for thirty minutes. The soundness of the pancake shall be checked at the end of the thirty minutes by passing the probe of a leak detector over the full outer surface of the pancake. This detector will be a mass spectrographic device tuned to the test gas, or similar system. Impregnation shall not be undertaken on any pancake exhibiting evidence of leakage.

### *2.9.6 Pancake insulation and impregnation*

2.9.6.1.) The pancakes shall be vacuum impregnated. The use of a mould is considered to be essential, and such a mould must apply direct contact pressure to as great a surface area of the pancake as can be achieved.

2.9.6.2.) Impregnation and curing shall be preceded by oven drying of a pancake and degassing of the resin, pancake and mould. Use of an open mould for the impregnation operation is preferred.

2.9.6.3.) The thickness of unreinforced resin on the surface of a finished pancake must not exceed 0.5 mm.

2.9.6.4.) After completion, the resin on the pancakes must be fully transparent, with no colouriser or additive that would limit observation of the copper turns used within the resin system. No paint or other external coating will be allowed.

2.9.6.5.) No pancake shall be repaired after its initial impregnation without the written approval of CELLS.

### *2.9.7 Terminations of the coil*

2.9.7.1.) The magnets shall be designed with all mechanical services (i.e. cooling water manifolds) and all electrical connections (including power terminals) on the inside of the ring at the downstream end of the bending magnet.

2.9.7.2.) Magnets shall be supplied with an inlet manifold and outlet manifold mounted on the mechanical services panel. The manifold shall be manufactured from metric stainless steel tube, grade 316, suitable for connection to the supply and return water system via a single Swagelok compression fitting onto each manifold. The manifold pipe will be mounted vertically on the mechanical services panel and the connection point will be at the bottom end of the tube.

2.9.7.3.) Magnets shall be equipped with a water flow controller Eletta type at the outlet manifold. All part of this controller in contact with water will be in brass with Canigen coating. The Eletta switch will include a witness window for visual flow indication.

2.9.7.4.) The water connections on the coils shall be manufactured from phosphor bronze, and shall be attached to the coil by a silver brazing technique to give a system which is unaffected by demineralised water.

2.9.7.5.) The electrical connections for power terminals shall be mounted on the electrical and mechanical services panel which shall be rigidly supported on the magnet. These terminals shall be suitable to receive the incoming supply cable connectors, and will be designed to withstand a maximum force of 50 kg exerted by the incoming cables on the terminals. The connection between the coil terminals and the services panel shall be the responsibility of the manufacturer.

2.9.7.6.) The manufacturer shall, during initial design, avoid locating any organic based material in the median plane of the magnet centre where it would be subject to long-term radiation damage from the beam. Coil water connections to the manifolds shall be well above and below the beam-line. Where water conduits cross the beam-line horizontal plane, pipe work shall be metallic.

2.9.7.7.) The coil terminals, the connection posts and all metallic parts connected to them will be protected against accidental contact by an insulating, transparent cover, which can only be removed by the use of tools; tenderers' proposals for this cover shall be described in the offer.

2.9.7.8.) The water connection between the coil terminals and the manifold shall be of non-conducting tube, having suitable mechanical properties, and suitable for use in a high radiation environment. Tenderers shall indicate in their quotation, the type of material that they propose for these tubes.

2.9.7.9.) The cabling of the over-temperature interlock switches shall be part of the contract. The two terminals of each switch shall be mounted on the electrical services board as specified.

2.9.7.10.) A single terminal connection post able to receive a 10 mm2 cable shall be provided for earthing the yoke. The manufacturer shall ensure that there is adequate electrical connection between the yoke, manifolds and other components so that all the exposed metallic parts of the magnet are safely earthed by this terminal post.

*2.9.8 Coils testing*

2.9.8.1.) The brazed termination shall be tested at a pressure of 60 bars for 10 minutes.

2.9.8.2.) During the total immersion of each coil in water the conductor shall be pressurised at 30 bar with water and sealed. The pressure shall be recorded, and any drop of pressure larger than 2 % during the 24 hours period shall result in rejection of the coil.

2.9.8.3.) The water flow for each of the water channels in a coil shall be separately measured with a pressure differential across the channel of 7 bar. The flow rate shall not be less than the flow rate, as calculated by the manufacturer and communicated to CELLS during the design phase, on which the coil thermal calculations are based. The flow in any coil shall also be in the range of ±10% of the mean for all coil flow measurements.

2.9.8.4.) The electrical resistance of all coils shall be measured with a DC bridge. The values shall be corrected to 23˚C, and must be within ±1% of the mean value for all coils.

2.9.8.5.) Each coil shall be immersed in tap water at ambient temperature, but with the terminals exposed above the water level. Any other part of the coil body not then completely immersed shall be covered with wet cloths the ends of which are in contact with the water. The following test sequence shall then be carried out:

a) Record insulation resistance between coil terminals and water bath, using minimum voltage of kV. Insulation resistance shall be above 50 MΩ.

b) Apply direct voltage of 5 kV between coil terminals and water bath for one minute, and record the leakage current.

c) Repeat measurement as in (a).

Any coil exhibiting evidence of breakdown or significant changes of insulation resistance during these tests shall be rejected.

2.9.8.6.) After completion of the tests in 2.9.8.5 each coil shall be energised until the coil temperature increases to 60 ˚C, as measured by the change in electrical resistance. During this period water shall be sealed within the conductor by means of a valve. On attaining the required temperature the current shall be interrupted and water at room temperature allowed to flow through the coil until the conductor again assumes the ambient temperature, as measured by the conductor resistance. The valve shall then again be closed and the foregoing cycle repeated fifty (50) times. The manufacturer may wish to undertake this procedure on a number of coils simultaneously. This test shall be performed in each one of the coils for the pre-series bending magnet and for one in each 5 coils for the series production.

2.9.8.7.) On completion of the thermal cycling the insulation tests described in shall be repeated, and significant changes of insulation resistance or breakdown characteristics shall again be sufficient reason for rejection of a coil. Any coil exhibiting evidence of cracking or delamination shall also be rejected.

2.9.8.8.) Immediately after the test described in 2.9.8.7 the coil shall be tested by using it as the secondary winding of a transformer. A maximum voltage of 2 kV RMS shall be induced across the coil terminations for a period of one minute, and the corresponding primary current recorded. Any indication of short-circuiting between turns shall result in rejection of the coil.

## 2.10 MECHANICAL AND ELECTRICAL TESTS ON COMPLETE MAGNETS

### 2.10.1 Mechanical and electrical tests on complete pre-series magnet

2.10.1.1.) After the magnets have been assembled with coils and cooling hoses, the complete assembly will be measured to ensure that it complies with the dimensional tolerances as specified.

2.10.1.2.) All the dimensions will be checked according to the appropriate magnet assembly drawing.

2.10.1.3.) The measuring techniques will be specifically designed to check each and all of the dimensions and dimensional tolerances defined in the appropriate assembly drawing and in this specification. These will be subject to CELLS approval. Manufacturers are requested to give details of their proposed assembly measurement techniques in their tenders.

2.10.1.4.) As an essential part of these tests and the subsequent production checks, the manufacturer shall develop and manufacture an instrument that is capable of making precision measurements of the gap region. This instrumentation is required to have the following features:

a) A measurement sensitivity and reproducibility equal to or better than ±10 μm;

b) A monitorable electrical output, allowing gap dimensions to be continuously measured as the gauge traverses through the magnet;

c) Accuracy and sensitivity unaffected by magnetic fields, i.e. to be capable of performing the required measurements with the bending magnet powered.

2.10.1.5.) A direct voltage of 5 kV shall be applied between the terminals of each coil and its magnet yoke for one minute. Any coil showing evidence of breakdown, indicated by a leakage resistance of less than 50 MΩ, shall be rejected.

2.10.1.6) A maximum d.c. operating current test with a coil excitation of 530 A shall be carried out for a period of at least two hours, with cooling water circuits set to provide a differential pressure not greater than 7 bar. During this test, the water inlet and outlet temperature shall be monitored, and the temperature of coil surfaces and all coil interconnections and terminals checked with contact thermometers. Results shall be judged with respect to the appropriate magnet thermal specifications. Any coil showing evidence of overheating, local hot spots or other faults during this period shall be rejected. This test can be performed at the commencement of the magnet measuring sequence, with the magnet connected to a power supply and cooling water.

2.10.1.7.) The manufacturer shall demonstrate the operating efficiency of the over-temperature switches by raising the temperature of each coil to the value (60 °C) at which the switches are guaranteed to operate. The technique that is to be used for the necessary overheating shall be agreed with CELLS. Because of the danger of damage to the coil, CELLS strongly prefers a method involving the external heating of the circulating water. Details of the proposed method shall be included in the tender.

2.10.1.8.) The manufacturer shall demonstrate the operating efficiency of the water flow switch by restricting slowly the supply of water with the magnet unpowered.

### 2.10.2 Mechanical and electrical tests on complete production magnets

2.10.2.1.) After the magnets have been assembled with coils, cooling hoses and mounting brackets fitted, the complete assembly will be measured to ensure that it complies with the dimensional tolerances as specified.

2.10.2.2.) All the dimensions will be checked according to the appropriate magnet assembly drawing.

2.10.2.3.) A direct voltage of 5 kV shall be applied between the terminals of each coil and its magnet yoke for one minute. Any coil showing evidence of breakdown, indicated by a leakage resistance of less than 50 MΩ, shall be rejected.

2.10.2.4.) A maximum d.c. operating current test with a coil excitation of 530 A shall be carried out for a period of at least two hours, with cooling water circuits set to provide a differential pressure not greater than 7 bar. During this test, the water inlet and outlet temperature shall be monitored, and the temperature of coil surfaces and all coil interconnections and terminals checked with contact thermometers. Results shall be judged with respect to the appropriate magnet thermal specifications. Any coil showing evidence of overheating, local hot spots or other faults during this period shall be rejected. This test can be performed at the commencement of the magnet measuring sequence, with the magnet connected to a power supply and cooling water.

*2.10.3    Acceptance tests after delivery*

2.10.3.1.) After delivery, the bending magnets shall be visually inspected for mechanical damage suffered in transit. Any such damage shall be reported to the manufacturer. Possible repair shall be subject to agreement with CELLS. Where the damage has resulted in alteration to the magnet iron geometry or to the soundness or shape of coil conductor, insulation or terminals, the magnet shall be rejected.

2.10.3.4.) Electrical tests shall be carried out by staff of CELLS after delivery. The manufacturer has the right to be represented during these tests but shall notify CELLS in writing if this right is to be exercised.

2.10.3.5.) A direct voltage of 5 kV will be applied between the terminals of each coil and its magnet yoke for one minute. Any coil showing evidence of breakdown, indicated by a leakage resistance of less than 50 MΩ, shall be rejected.

2.10.3.6.) The magnet will be energised at the maximum current of 530 A for a period of at least two hours. Any coil showing evidence of breakdown, local hot spots or other faults during this period shall be rejected.

## 2.11 PACKING AND TRANSPORTATION

*2.11.1    Packing*

2.11.1.1.) The contractor will submit to CELLS a solution for the packing. This packing will have to use the classical handling tools. The magnet packed shall be protected against the elements, the projections and the breaks during transportation and storage.

2.11.1.2.) The packing of each magnet shall be dust proof, water proof and will have to protect the steel parts against the oxidation. Moreover, the pieces shall be protected against the strain, impacts and rubbing which can damage their surfaces.

2.11.1.3.) A particular protection shall be required for brittle parts (reference surfaces, electrical connections and coils).

2.11.1.4.) The coils shall be rinsed, dried and sealed before expedition in order to avoid any frost.

*2.11.2    Transportation*

2.11.2.1.) The contractor will include in his tender the transport from the factory:

a) To the site where CELLS will control and measure the magnets. This place will not be necessarily in the CELLS site.

b) Or to the storage area that CELLS will indicate. This storage area will be at CELLS site or nearby.

2.11.2.2.) The contractor keeps the responsibility of the bending magnets until the delivery to one of the sites listed above. CELLS will supply the local handling tools.

2.11.2.3.) The transfer of risks shall take place when the load is laid down on the ground.

**Acknowledgement**

The author wants to acknowledge the help and contribution of different colleagues: Montse Pont (CELLS/ALBA), Davide Tommasini (CERN), Neil Marks (Crockcroft Institute), and G. Moritz (GSI). The author is indebted to Jordi Marcos (ALBA/CELLS) for his excellent assistance in preparing the manuscript.

## Appendix A: Design criteria for different types of magnets

In this appendix we recall the main points concerning the design of iron-dominated magnets (bending, quadrupole, and sextupole).

### A.1 Design criteria of a dipole magnet

A dipole or bending magnet has two poles generating a constant field which steers the particle beam. The purpose of the complete set of bending magnets in a circular accelerator is to bend the beam by exactly 360 degrees. Some examples of bending magnets are shown in Fig. A1.

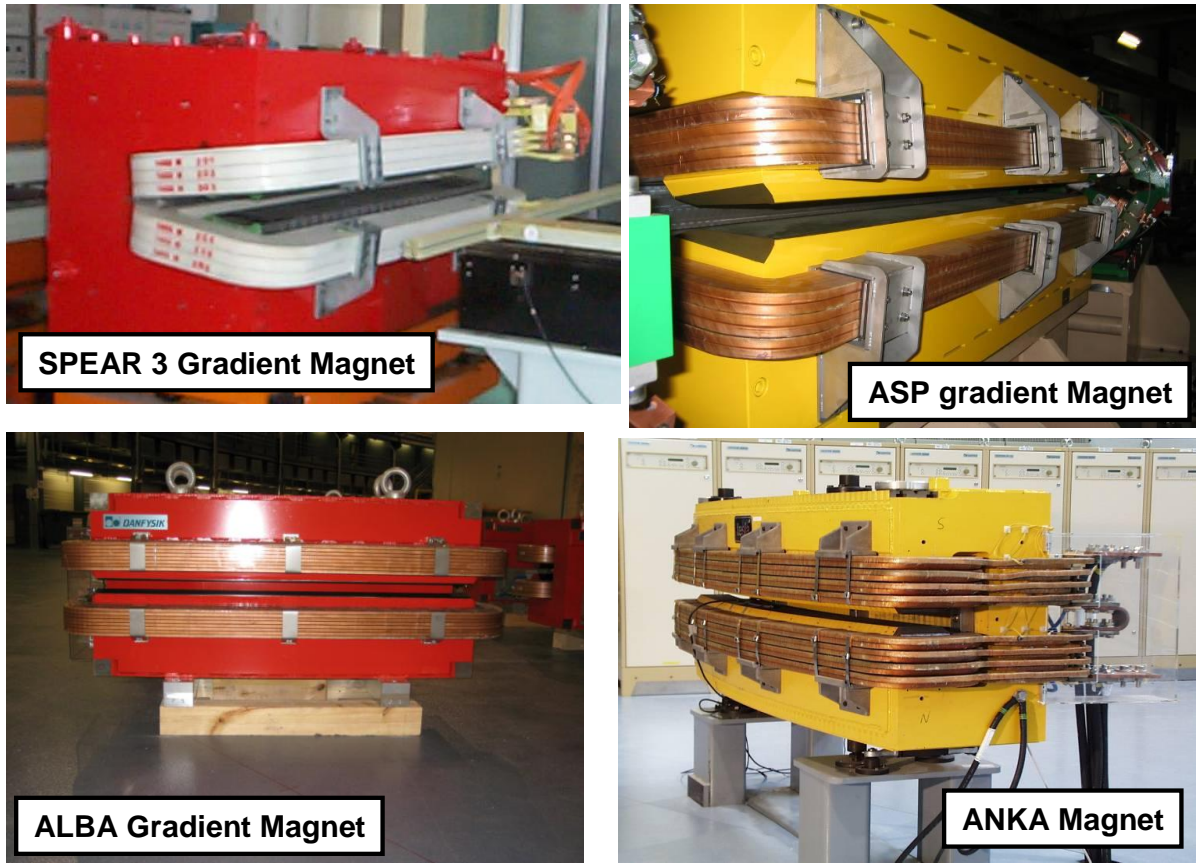

**Fig. A1:** Examples of dipole magnets for different synchrotron light source facilities

The gap ($g = 2\,h$) in a normal bending magnet is constant:

$$h(x) = \text{const.} = {gap}/{2} \tag{A1}$$

At the ends of the pole there are shims introduced to optimize the 'good field region area' (see Fig. A2).

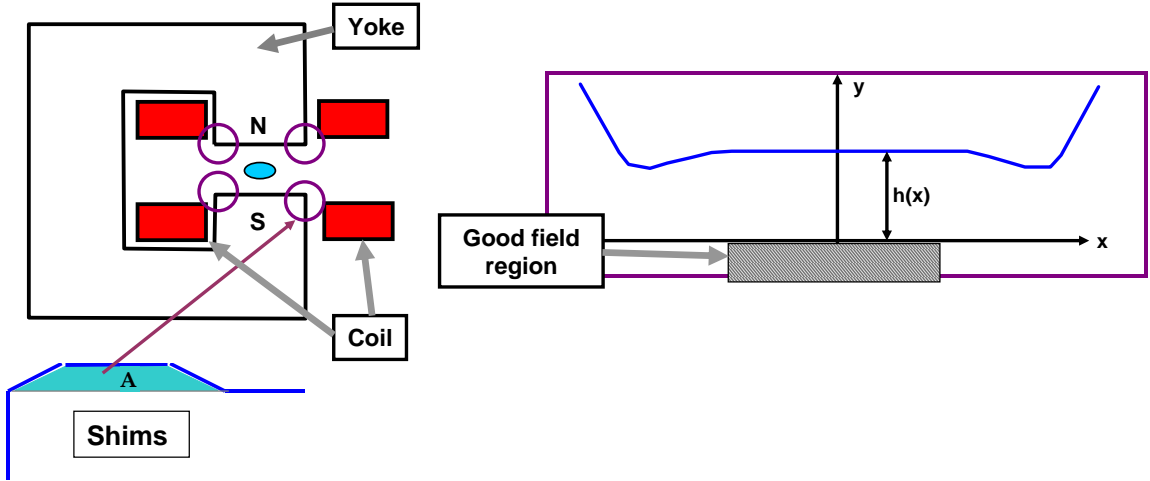

**Fig. A2:** *Left:* Schematic cross section of a bending magnet and detail of the end of the pole with a shim. *Right:* Pole profile with an indication of the good field region of the magnet.

The size of a dipole is determined by the dimensions of the 'good field region' required by accelerator physicists. For example, in a typical bending magnet for a synchrotron light source a good field region of roughly ±15 to ±20 mm is required. In heavy-ion accelerators or other machines the dimensions of the good field region can be completely different (usually much larger).

To optimize the pole profile the so-called shims are used at the end of the poles (as already shown in Fig. A2). The contour of the shims has to be determined with a 2D (Poisson) or 3D (TOSCA) magnetic field simulation code.

The main characteristic of a bending magnet is to deflect the particle beam by an angle $\varphi$. The deflection angle is proportional to the integrated field of the magnet, and the change $\Delta\varphi$ of the deflection angle is given by the change of the magnetic field ($\Delta B$) over the length of the magnet, $\Delta s$. For the correction of $\Delta\varphi$ so-called correctors are introduced into the lattice.

$$\varphi = \frac{\int B \, ds}{B_0 \rho_0}$$
$$\Delta\varphi = \frac{\Delta B \, \Delta s}{B_0 \rho_0}$$
(A2)

The bending angle of the Storage Ring bending magnets of ALBA is 11.25 degrees (or 196.35 mrad). The maximum allowed change of $\Delta\varphi$ is determined by the acceptance of the machine. For instance, in the case of ALBA $\Delta\varphi$ should be smaller than 0.3 mrad, which means that $\Delta\varphi/\varphi \leq 0.0015$ or $1.5 \times 10^{-3}$. A typical requirement is that $\Delta\varphi/\varphi \leq 0.001$ or $1 \times 10^{-3}$. That is to say, with a constant field $B_0$ everywhere in the bending magnet the tolerance of the length has to be better than $\pm 0.5 \times 10^{-3}$. For a 1 metre-long bending magnet the resulting tolerance is $\pm 0.5$ mm, which can be achieved.

The change of the deflection angle $\Delta\varphi$ is proportional to the change of the flux density $\Delta B$ and to the change of the gap $\Delta g$.

$$\frac{\Delta B}{B_0} = \frac{\Delta\phi}{\phi} = \frac{\Delta g}{g}.$$
(A3)

With a gap of roughly 40 mm and the requirement of $\Delta\varphi/\varphi \leq 1 \times 10^{-3}$, the change of the gap has to be smaller than 40 µm. As a consequence the gap height has to be very accurate, in the range of ±20 µm.

*A.1.1    Pole profiles in a bending magnet*

In some cases there is a gradient ($G$) in the bending magnet so as to have a so-called combined function bending magnet. To reach this a slope has to be built into the gap of the bending magnet. The change of the slope is roughly given by the product of the gradient and the gap divided by the field.

$$\frac{\Delta y}{\Delta x} = \frac{G(g/2)}{B_0} = \frac{(g/2)}{X_0} \ . \tag{A4}$$

For the Storage Ring of ALBA ($G = 5.6$ T/m, $g = 40$ mm, $B_0 = 1.42$ T) the resulting slope is roughly 78.9 mrad or, for a distance of $x = 25$ mm away from the centre, a value of 2.5 mm. With the required tolerance of $1 \times 10^{-3}$, the change of the gap should have an accuracy of roughly 2.5 µm. This is pretty tight and a fairly impossible tolerance to be fulfilled.

More precisely, if a gradient has to be introduced into the bending magnet the gap needs a hyperbolic shape (Fig. A3). The change of the half-gap height with the horizontal position is given by

$$h(x) = \frac{h(0)}{1 - \frac{xG}{B_0}} = \frac{h(0)}{1 - \frac{x}{X_0}} \quad \text{(hyperbolic approach)} \tag{A5}$$

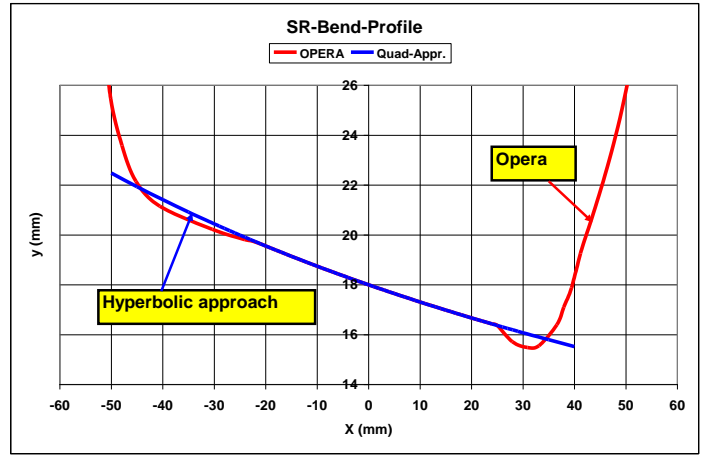

**Fig. A3:** Pole profile of the gradient bending magnet for the Storage Ring of ALBA determined using a hyperbolic approach (blue line) or using OPERA3D field simulation code (red line)

If a sextupole component $B''$ has to be introduced into the bending magnet the gap needs a quadratic shape. In this case the change of the half-gap is given by

$$h(x) = h(0)\left(1 + \frac{(x/X_0)^2}{1 - (x/X_0)^2}\right) \quad \text{with} \quad X_0 = \sqrt{\frac{2B_0}{B''}} \ . \tag{A6}$$

For example, for a sextupole component of $B'' = 40$ T/m, the half-gap has to be increased (at a transverse distance of 30 mm) by 0.257 mm. The pole profile can be machined with an accuracy of roughly ± 20 µm, which means that the sextupole component in a dipole does not have a high accuracy (5 to 15%).

## A.2 Design criteria of a quadrupole magnet

A quadrupole magnet has four poles. The field varies linearly with the distance from the magnet centre. It focuses the beam within one plane while defocusing the beam within the orthogonal plane.

The field of the quadrupole has to be proportional to the distance from the centre (along either $x$ or $y$). The excitation in general is given by

$$B_0 = \frac{\mu_0 N I}{g} \quad \text{or} \quad B(x) = \mu_0 N I \, x \,. \tag{A7}$$

This means that the pole profile $g(x)$ of a quadrupole has to be a hyperbola, as shown in Fig. A4.

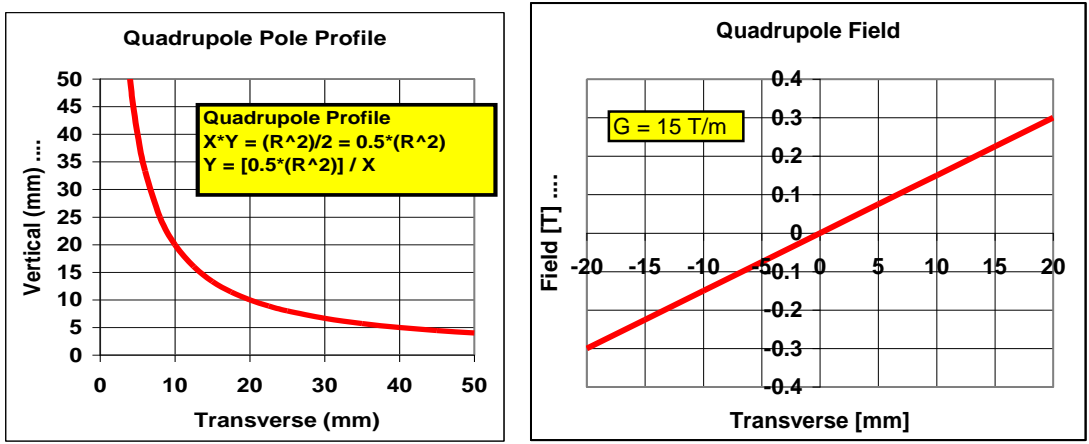

**Fig. A4:** Pole profile (left) and dependence of the magnetic field along the horizontal direction (right) for a quadrupole magnet

Some examples of quadrupole magnets are shown in Fig. A5.

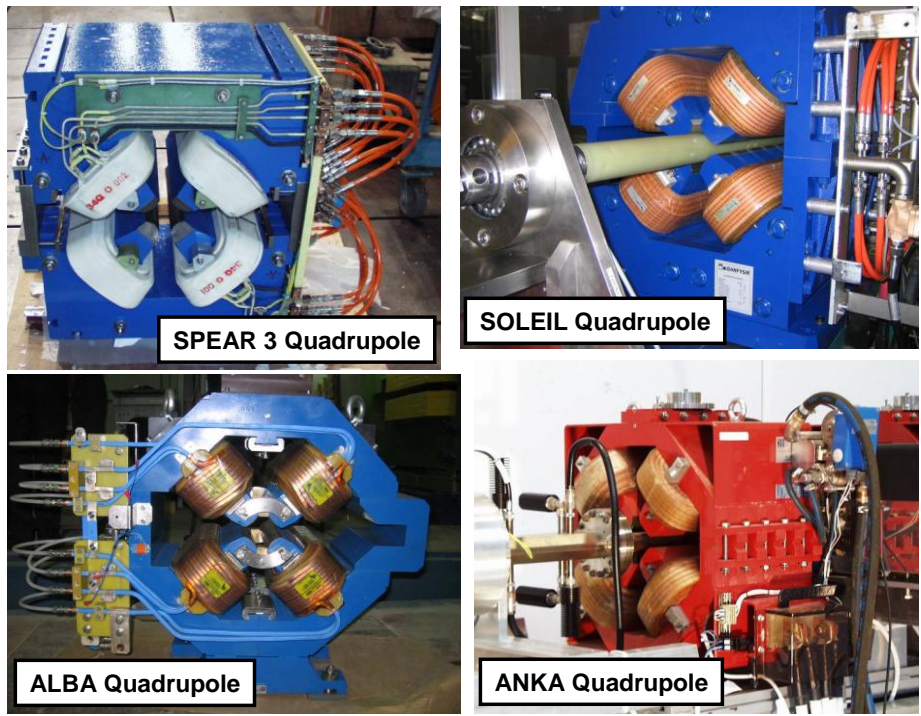

**Fig. A5:** Examples of quadrupole magnets for different synchrotron light source facilities

By cutting the pole profile in order to have space for the introduction of the coils, the field distribution will be disturbed and higher order systematic multipoles will be introduced. In the case of fully symmetric magnets, the allowed systematic multipoles depending on the number of poles of the magnet are listed in Table A1.

**Table A1:** Allowed systematic multipoles depending on the fundamental geometry of the magnet, for fully symmetric magnets

| Fundamental geometry | Systematic multipoles |
| --- | --- |
| Dipole, $n=1$ | $n = 3, 5, 7…$ (6-pole, 10-pole, 14-pole…) |
| Quadrupole, $n=2$ | $n = 6, 10, 14…$ (12-pole, 20-pole…) |
| Sextupole, $n=3$ | $n = 9, 15, 21…$ (18-pole, 30-pole…) |
| Octupole, $n=4$ | $n = 12, 20, 28…$ (24-pole, 40-pole…) |

Assembly errors introduce higher multipoles as well. In this case they are called random multipoles. In the case of a quadrupole magnet some of the possible assembly errors are illustrated in Fig. A6.

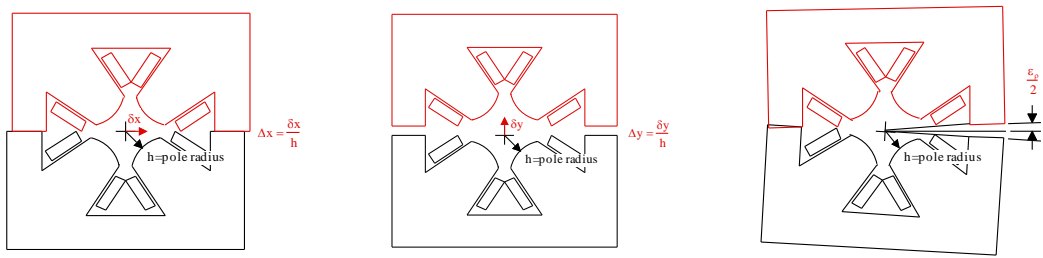

**Fig. A6:** Some of the possible assembly errors for a quadrupole magnet

Figure A7 shows two possible ways of segmenting the laminations of a quadrupole. In approach (a), each segment can be assembled with errors with three kinematic motions, x, y and e (rotation). Thus, combining the possible errors of the three segments with respect to the datum segment, the core assembly can be assembled with errors with $3 \times 3 \times 3 = 27$ degrees of freedom. In comparison, approach (b) has the advantage that the two core halves can be assembled kinematically with only three degrees of freedom for assembly errors. Thus assembly errors are more easily measured and controlled.

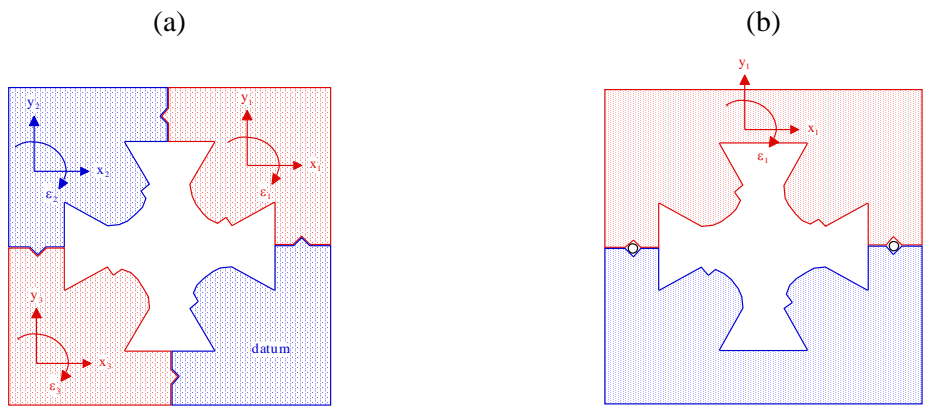

**Fig. A7:** Two possible ways of segmenting the laminations of a quadrupole magnet

## A.2.1 Higher multipoles in quadrupoles

The geometrical parameters involved in the design of a quadrupole magnet and determining its content of higher multipoles are illustrated in Fig. A8. The effect of introducing different modifications of these parameters on the content of higher multipoles is listed in Table A2. It can be concluded that to avoid higher multipoles the manufacturing of the magnets has to be very accurate (A = B and a = b = c = d).

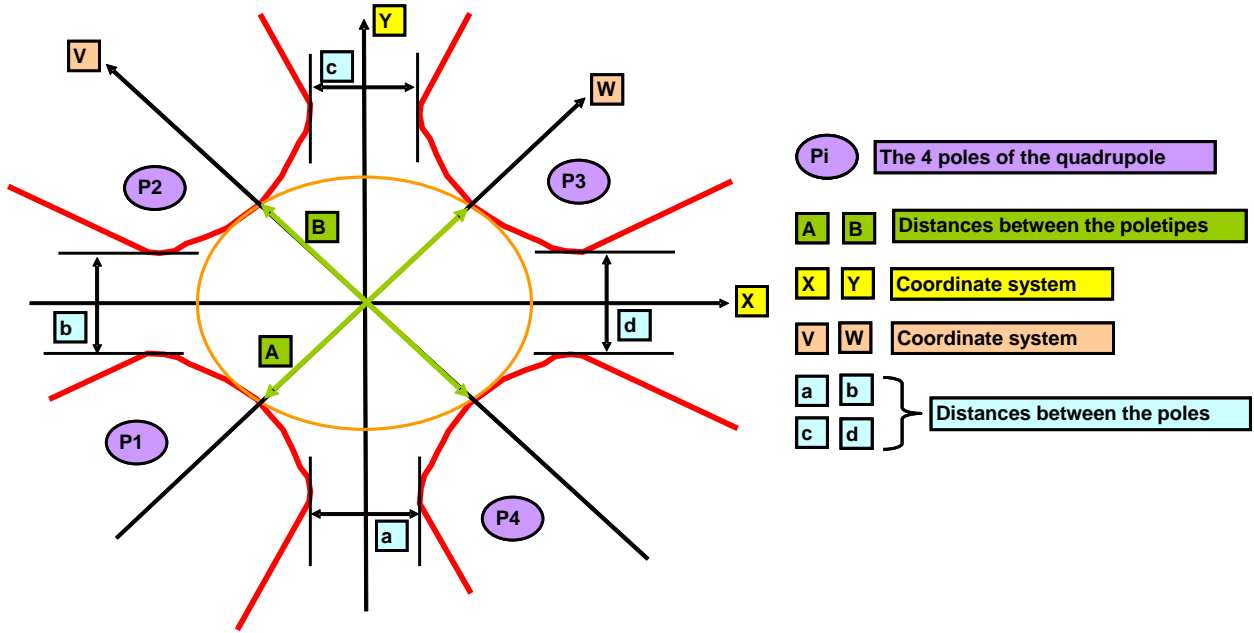

**Fig. A8:** Geometrical parameters involved in the design of a quadrupole

**Table A2:** Introduced higher order multipoles for different modifications of the geometrical parameters of a quadrupole magnet as defined in Fig. A8

|    | Geometrical change | Allowed multipoles |
|----|-------------------|--------------------|
| 0) | Fully symmetric quadrupole; A=B and a=b=c=d and poles are truncated | $n=6$ |
| 1) | P1 in neg. W | $n=3, n=4, n=5, n=6, n=10$ |
| 2) | P4 in neg. V | $n=3, n=4, n=5, n=6, n=10$ |
| 3) | P3 in neg. W | $n=3, n=4, n=5, n=6, n=10$ |
| 4) | P2 in pos. Y | $n=3, n=4, n=5$ |
| 5) | P2 in neg. Y | $n=3, n=4, n=5$ |
| 6) | P2 in neg. V and P4 in pos. V | $n=4, n=6, n=8, n=10$ |
| 7) | P2 in pos. X | $n=3, n=4, n=5$ |
| 8) | P2 in neg. X | $n=3, n=4, n=5$ |
| 9) | A and B are increased by the same amount | $n=6, n=10$ |
| 10) | A and B are decreased by the same amount | $n=6, n=10$ |
| 11) | A and B not equal | $n=4, n=6, n=8, n=10$ |

## A.3 Design criteria of a sextupole magnet

A sextupole magnet has six poles. The field varies quadratically with the distance from the magnet centre, as shown in Fig. A9. The purpose of sextupoles is to affect the beam at the edges, much like an optical lens which corrects chromatic aberration. Sextupoles are needed for the compensation of the chromaticity of the machine or, in other words, in order to make the focusing of the machine independent of the energy of the beam in a small range. Note that the sextupole also steers the beam along the 60 and 120 degree lines.

Several examples of sextupole magnets are shown in Fig. A10.

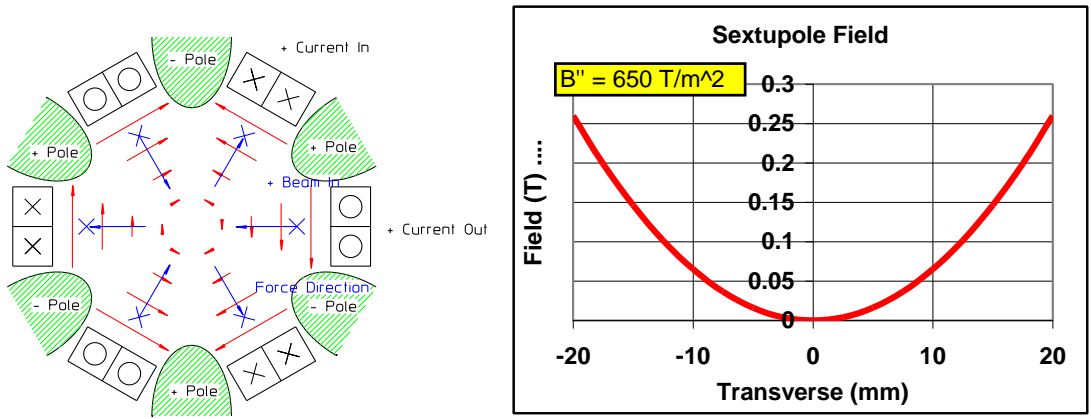

**Fig. A9:** Dependence of the magnetic field along the horizontal direction for a sextupole magnet

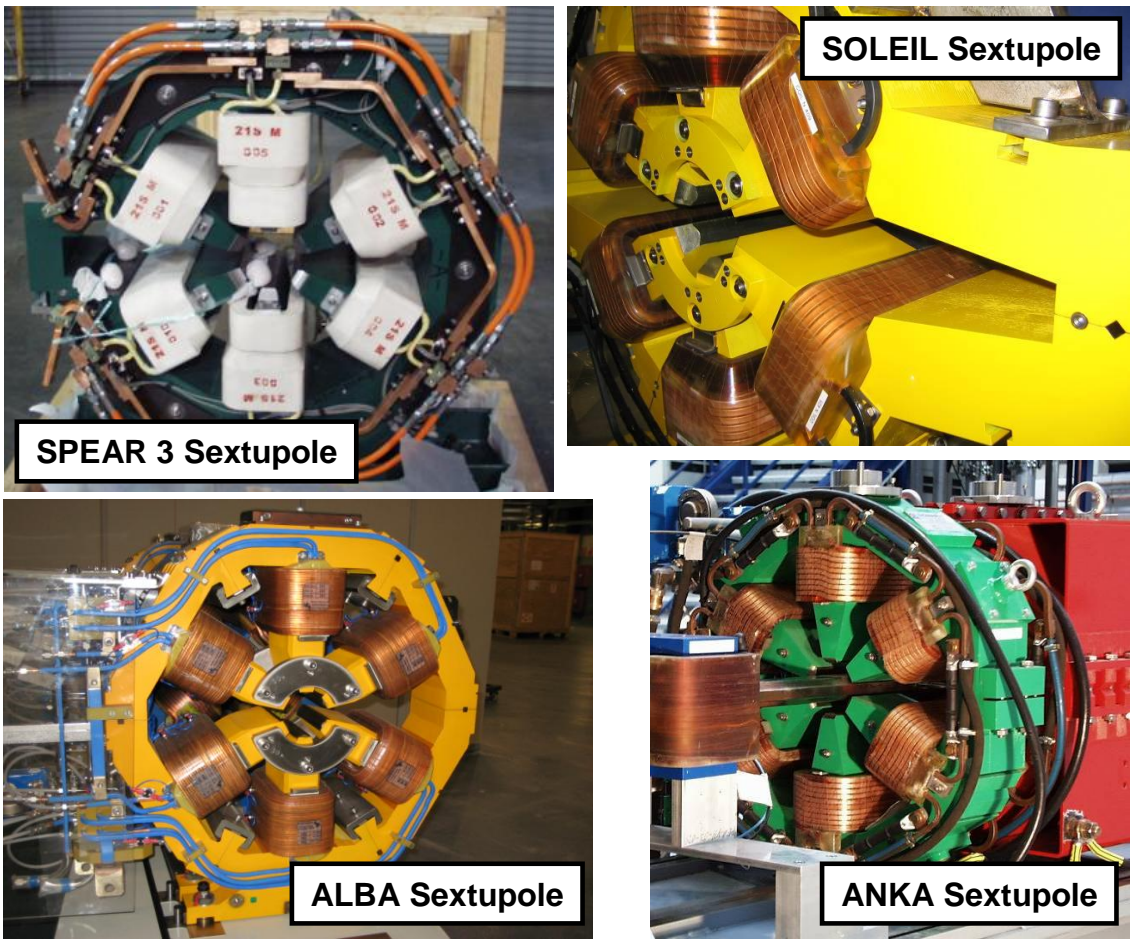

**Fig. A10:** Examples of sextupole magnets for several synchrotron light sources

## A.4 Specification of coils

The current and the number of windings of a coil determine the excitation of a magnet through the following expression:

$$NI = \frac{B_0}{\mu_0}\left(g + \frac{\ell_{Fe}}{\mu_r}\right) \approx \frac{B_0 \, g}{\mu_0}.$$  (A8)

For ALBA Storage Ring bending magnets with a gap of 36 mm and a flux density of 1.42 T the required excitation is

$$NI = 40\,680 \text{ Amp-turns}.$$

This number can be attained either by using a larger number of turns or by increasing the current.

The standard design for magnet coils is rectangular copper (or aluminium) conductor, with a cooling water tube inside, as shown Fig. A11. Either glass cloth or epoxy resin can be used as insulation.

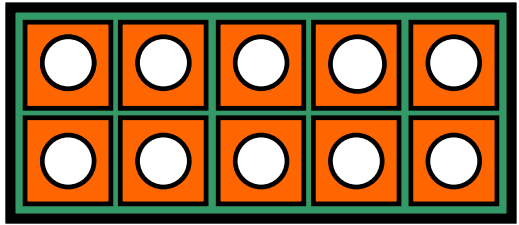

**Fig. A11:** Cross section sketch of the standard design for magnet coils. The shown coil corresponds to 1 pancake with 10 turns, arranged in 2 layers.

Once the required number of Amp-turns (*NI*) has been determined, there is still some freedom to choose the total copper area ($A_{copper}$) and the number of turns (*N*) of the coils. The choice is made based on economic criteria taking into account the current density (*j*) circulating through the windings of the coil, given by

$$j = \frac{NI}{A_{copper}}.$$  (A9)

The arrangement of the windings within the cross section of the coil has to be decided taking into account the space requirements of the machine. Usually, it is desirable to reduce the space between the components of the machine as much as possible. A small number of layers needs more space in the longitudinal directions, whilst a high number of layers needs more space for the magnets in the vertical direction. Therefore a compromise that will depend on the characteristics of the machine has to be reached.

## A.4 Other design criteria

For the design of the magnets one has to take further into account:

- The requirements of the vacuum system.
- The requirements (fixing) of the diagnostics elements.
- The requirements of the front ends.
- Fixing of the magnets on the girders/supports.

**Appendix B: Sequence for magnet production**

A typical sequence for the production of the magnets of an accelerator includes the following steps:

1) The requirements of the magnets (field, length, gradient, sextupole field, good field region, etc.) are established by the Beam Dynamics Group.

2) A basic design of the magnets is prepared by the Magnet Group (pole profile, H- or C-type of magnet, cross section of the lamination, coil design, supports, etc.), including the required space needed for vacuum, diagnostics, cooling, etc.

3) Steps 1) and 2) have to be made according to the space requirements for the vacuum, diagnostics, etc., and hence some iterations are needed before finishing the basic design.

4) At this point one of the following decisions has to be taken:
    a) Will the detailed design be made in-house by the magnet group/engineering group?
    b) Will the detailed design be made by the manufacturer?

    In most cases the detailed design is made by the manufacturer. The reason for this is that the manufacturer has to prepare the production drawings anyway and therefore he should also make the detailed design.

5) The fulfillment of the specifications can only be checked by means of magnetic measurements. Therefore it has to be decided by whom and where the magnetic measurements will be carried out. This has to be written down in the Technical Specifications for the Call for Tender.

6) Going out for the Call for Tender.

7) Evaluation of the different offers, including some discussions with the bidders. Choice of the manufacturer.

8) Signature of the contract.

9) Preparation of the detailed design by the manufacturer.

10) Acceptance of the detailed design (coils, manifolds, etc.) by the customer.

11) Preparation of the manufacturing drawings.

12) Production of stamping tools.

13) Procurement of raw materials: steel, copper, etc.

14) Acceptance test of laminations.

15) Production of prototype or pre-series magnet.

16) Magnetic measurement of prototype magnet.

17) Determination of end chamfer, acceptance of prototype and agreement of modifications for the series production.

18) Series production.

19) Mechanical test of the yokes.

20) Magnetic and electrical test of the coils.

21) Magnetic measurements of the series magnets.

22) Acceptance of the magnets.

All the previous points have to be addressed in the Technical Specifications document.